\RequirePackage{fix-cm}
\documentclass[smallextended]{svjour3}
\smartqed
\usepackage{graphicx}
\usepackage{booktabs}
\usepackage{pgfplots}
\usepackage{makecell}
\usepackage{subfigure}
\usepackage[hidelinks]{hyperref}
\usepackage[numbers]{natbib}
\usepackage{tcolorbox}
\usepgfplotslibrary{statistics}
\pgfplotsset{compat=1.14}
\begin{document}
\begin{sloppy}
\title{Categorizing the Content of GitHub README Files}
\author{Gede Artha Azriadi Prana \and Christoph Treude \and Ferdian Thung \and Thushari Atapattu \and David Lo}
\institute{
Gede Artha Azriadi Prana \and Ferdian Thung \and David Lo \at Singapore Management University\\
\email{arthaprana.2016@phdis.smu.edu.sg, ferdiant.2013@phdis.smu.edu.sg, davidlo@smu.edu.sg}
\and
Christoph Treude \and Thushari Atapattu \at University of Adelaide\\
\email{christoph.treude@adelaide.edu.au, thushari.atapattu@adelaide.edu.au}
}

\date{Received: date / Accepted: date}
\maketitle
\begin{abstract}
README files play an essential role in shaping a developer's first impression of a software repository and in documenting the software project that the repository hosts. Yet, we lack a systematic understanding of the content of a typical README file as well as tools that can process these files automatically.
To close this gap, we conduct a qualitative study involving the manual annotation of 4,226 README file sections from 393 randomly sampled GitHub repositories and we design and evaluate a classifier and a set of features that can categorize these sections automatically.
We find that information discussing the `What' and `How' of a repository is very common, while many README files lack information regarding the purpose and status of a repository. Our multi-label classifier which can predict eight different categories achieves an F1 score of 0.746.
To evaluate the usefulness of the classification, we used the automatically determined classes to label sections in GitHub README files using badges and showed files with and without these badges to twenty software professionals. The majority of participants perceived the automated labeling of sections based on our classifier to ease information discovery. 
This work enables the owners of software repositories to improve the quality of their documentation and it has the potential to make it easier for the software development community to discover relevant information in GitHub README files.
\keywords{GitHub README files \and Classification \and Documentation}
\end{abstract}

\section{Introduction and Motivation}
The \texttt{README.md} file for a repository on GitHub is often the first project document that a developer will see when they encounter a new project. This first impression is crucial, as Fogel~\cite{Fogel2005} states: ``The very first thing a visitor learns about a project is what its home page looks like. [...] This is the first piece of information your project puts out, and the impression it creates will carry over to the rest of the project by association.''

With more than 25 million active repositories at the end of 2017\footnote{\url{https://octoverse.github.com/}}, GitHub is the most popular version control repository and Internet hosting service for software projects. When setting up a new repository, GitHub prompts its users to initialize the repository with a \texttt{README.md} file which by default only contains the name of the repository and is displayed prominently on the homepage of the repository.

A recent blog post by Christiano Betta\footnote{\url{https://betta.io/blog/2017/02/07/developer-experience-github-readmes/}} compares the README files of four popular GitHub repositories and stipulates that these files should (1) inform developers about the project, (2) tell developers how to get started, (3) document common scenarios, and (4) provide links to further documentation and support channels. In its official documentation\footnote{\url{https://help.github.com/articles/about-readmes/}}, GitHub recommends that a README file should specify ``what the project does, why the project is useful, how users can get started with the project, where users can get help with your project, and who maintains and contributes to the project''. Brian Doll of GitHub claimed in a recent interview for IEEE Software that ``the projects with good README files tend to be the most used, too, which encourages good README writing behavior''~\cite{Begel2013}. 

In the research literature, GitHub README files have been used as a source for automatically extracting software build commands~\cite{Hassan2017}, developer skills~\cite{Greene2016, Hauff2015}, and requirements~\cite{Portugal2016}. Their content has also played a role in cataloguing and finding similar repositories~\cite{Sharma2017, Zhang2017} as well as in analyzing package dependency problems~\cite{Decan2016}. 

However, up to now and apart from some anecdotal data, little is known about the content of these README files. To address this gap, our first research question RQ1 asks, \textit{What is the content of GitHub README files?} Knowing the answer to this question would still require readers to read an entire file to understand whether it contains the information they are looking for. Therefore, our second research question RQ2 investigates, \textit{How accurately can we automatically classify the content of sections in GitHub README files?}. To understand a README file's most defining features, our third research question RQ3 asks, \textit{What value do different features add to the classifier?}. Finally, to evaluate the usefulness of the classification, our last research questions RQ4 investigates, \textit{Do developers perceive that the automated classification makes it easier to discover relevant information in GitHub README files?}

To answer our research questions, we report on a qualitative study of a statistically representative sample of 393 GitHub README files containing a total of 4,226 sections. Our conclusions regarding the frequency of section types generalize to the population of all GitHub README files with a confidence interval of 4.94 at a confidence level of 95\%. Our annotators and ourselves annotated each section with one or more codes from a coding schema that emerged during our initial analysis. This annotation provides the first large-scale empirical data on the content of GitHub README files. We find that information discussing the `What' and `How' of a repository is common while information on purpose and status is rare. These findings provide a point of reference for the content of README files that repository owners can use to meet the expectations of their readers as well as to better differentiate their work from others.

In addition to the annotation, we design a classifier and a set of features to predict categories of sections in the README files. This enables both quick labeling of the sections and subsequent discovery of relevant information. We evaluated the classifier's performance on the manually-annotated dataset, and identify the most useful features for distinguishing the different categories of sections. Our evaluation shows that the classifier achieves an F1 score of 0.746. Also, the most useful features are commonly related to some particular words, either due to their frequency or their unique appearance in sections' headings. In our survey to evaluate the usefulness of the classification, the majority of twenty software professionals perceived the automated labeling of sections based on our classifier to ease information discovery in GitHub README files.

We make the following contributions:

\begin{itemize}
    \item A qualitative study involving the manual annotation of the content of 4,226 sections from 393 randomly selected GitHub README files, establishing a point of reference for the content of a GitHub README file. We distinguish eight categories in the coding schema that emerged from our qualitative analysis (What, Why, How, When, Who, References, Contribution, and Other), and we report their respective frequencies and associations.
    \item We design and evaluate a classifier that categorizes README sections, based on the categories discovered in the annotation process.
    \item We design and conduct a survey to evaluate the usefulness of the classification by (i) using the automatically determined classes to label sections in GitHub README files using badges and (ii) showing files with and without these badges to twenty software professionals.
\end{itemize}

We describe background materials on GitHub and README files of repositories hosted there in Section~2. We describe our manual annotation methodology in Section~3 and the results of the annotation in Section~4. Section~5 introduces the classifier we built for sections of GitHub README files, which we evaluate in Section~6. We discuss the implications of our work in Section~7 and present the threats to validity associated with this work in Section~8. We review related work in Section~9 before we conclude in Section~10.

\section{Background}

GitHub is a code hosting platform for version control and collaboration.\footnote{\url{https://guides.github.com/activities/hello-world/}} Project artifacts on GitHub are hosted in repositories which can have many branches and are contributed to via commits. Issues and pull requests are the primary artifacts through which development work is managed and reviewed.

Due to GitHub's pricing model which regulates that public projects are always free\footnote{\url{https://github.com/open-source}}, GitHub has become the largest open source community in the world, hosting projects from hobby developers as well as organizations such as Adobe, Twitter, and Microsoft.\footnote{\url{https://github.com/collections/open-source-organizations}}

Each repository on GitHub can have a README file to ``tell other people why your project is useful, what they can do with your project, and how they can use it.''\footnote{\url{https://help.github.com/articles/about-readmes/}} README files on GitHub are written in GitHub Flavored Markdown, which offers special formatting for headers, emphasis, lists, images, links, and source code, among others.\footnote{\url{https://guides.github.com/features/mastering-markdown/}} Figure~\ref{fig:example} shows the README file of D3, a JavaScript library for visualizing data using web standards.\footnote{\url{https://github.com/d3/d3}} The example shows how headers, pictures, links, and code snippets in markdown files are represented by GitHub.

\begin{figure}
\centering
\includegraphics[width=\linewidth]{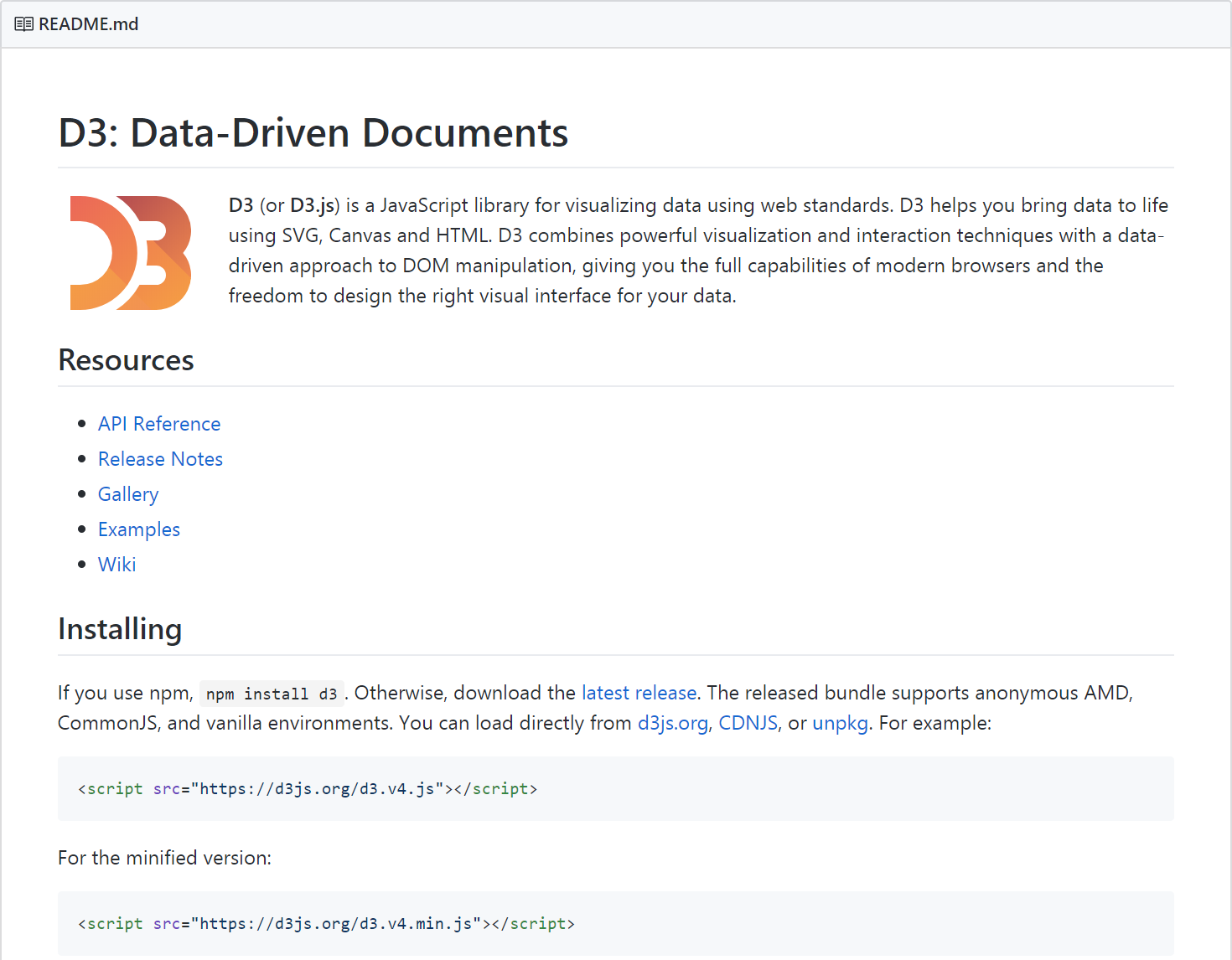}
\caption{An excerpt from D3's GitHub README file}
\label{fig:example}
\end{figure}

With 1 billion commits, 12.5 million active issues, and 47 million pull requests in the last 12 months, GitHub plays a major role in today's software development landscape.\footnote{\url{https://octoverse.github.com/}} In 2017, 25 million active repositories were competing for developers' attention, and README files are among the first documents that a developer sees when encountering a new repository.

To gain an understanding of readers' expectations about README files, in our survey to evaluate our classifier, we asked participants what content they expect to find in the README file of a GitHub repository and what single piece of information they would consider most important to be included. Twenty professionals answered our survey---we refer readers to Section~\ref{sec:survey} for details on survey design and participant demographics. Here, we summarize the responses we received regarding readers' expectations about the content of GitHub README files.

In response to the open-ended question ``What content do you expect to find in the README file of a GitHub repository?'', participants mentioned usage instructions (five participants), installation instructions (three participants), prerequisites (three participants), repository license (two participants), purpose of the repository and target audience (two participants), known bugs and trouble-shooting tips (two participants), coding style (one participant), contribution guidelines (one participant), change log (one participant), and screenshots (one participant). For example, one participant answered ``Information about the program, how to use it, parameters (if applicable), trouble-shooting tips (if applicable)'' and another indicated ``I expect to see how to install and run the program successfully''. Nine of the twenty participants provided generic answers, such as ``More technical information and guidance'' and ``updates''. 

In response to ``What single piece of information would you consider most important to be included in a GitHub README file?'', we also received twenty responses. Usage instructions (e.g., ``How to use the features or components of the repository'') and license information (e.g., ``With my job it's most important to know the licensing information'') were identified as most important by three participants each. Two participants indicated  known bugs and trouble-shooting tips as being most important, while the other participants mentioned a variety of types of information including target audience, coding style, contribution guidelines, testing information, prerequisites, screenshots and demos, and project type.

In this work, we study and classify the content of README files on GitHub to investigate the extent to which these expectations are met.

\section{Research Methodology}

In this section, we present our research questions and describe the methods for data collection and analysis.

\subsection{Research Questions}

Our work was guided by four research questions, which focus on categorizing the content of GitHub README files and on evaluating the performance and usefulness of our classifier:

\begin{description}
\item [RQ1] What is the content of GitHub README files?
\end{description}

Answers to this question will give insight to repository maintainers and users about what a typical README file looks like. This can serve as a guideline for repository owners who are trying to meet the expectations of their users, and it can also point to areas where owners can make their repositories stand out among other repositories.

\begin{description}
\item [RQ2] How accurately can we automatically classify the content of sections in GitHub README files?
\end{description}

Even after knowing what content is typically present in a GitHub README file, readers would still have to read an entire file to understand whether it contains the kind of information they are looking for. An accurate classifier that can automatically classify sections of GitHub README files would render this tedious and time-consuming step unnecessary. From a user perspective, an automated classifier would enable a more structured approach to searching and navigating GitHub README files.

\begin{description}
\item [RQ3] What value do different features add to the classifier?
\end{description}

Findings to our third research question will help practitioners and researchers understand the content of README files in more detail and shed light on their defining features. These findings can also be used in future work to further improve the classification.

\begin{description}
\item [RQ4] Do developers perceive that the automated classification makes it easier to discover relevant information in GitHub README files?
\end{description}

The goal of our last research question is to evaluate the usefulness of the automated classification of sections in GitHub README files. We use the automatically determined classes to label sections in unseen GitHub README files using badges, and we show GitHub README files with and without these labels to developers and capture their perceptions regarding the ease of discovering relevant information in these files.

\subsection{Data Collection}

To answer our research questions, we downloaded a sample of GitHub \texttt{README.md} files\footnote{We only consider \texttt{README.md} files in our work since these are the ones that GitHub initializes automatically. GitHub also supports further formats such as \texttt{README.rst}, but these are much less common and out of scope for this study.} by randomly selecting GitHub repositories until we had obtained a statistically representative sample of files that met our selection criteria. We excluded README files that contained very little content and README files from repositories that were not used for software development. We describe the details of this process in the following paragraphs.

To facilitate the random selection, we wrote a script that retrieves a random GitHub repository through the GitHub API using the \texttt{https://api.github.com/repositories?since=<number>} API call, where \texttt{$<$number$>$} is the repository ID and was replaced with a random number between 0 and 100,000,000, which was a large enough number to capture all possible repositories at the time of our data collection. We repeated this process until we had retrieved a sufficient number of repositories so that our final sample after filtering would be statistically representative. We excluded repositories that did not contain a README file in the default location.

\begin{table}
\centering
\caption{Number of repositories excluded from the sample}
\label{tab:readme-files}
\begin{tabular}{lr}
\toprule
Reason for Exclusion & Repositories \\
\midrule
Software, but small README file, i.e., $<$ 2 KB & 429 \\
Not software, but large enough README file & 127 \\
Not software and small README file & 196 \\
\midrule
README file not in English & 48 \\
\midrule
Number of repositories included in the sample & 393 \\
\midrule
Total number of repositories inspected & 1,193 \\
\bottomrule
\end{tabular}
\end{table}

Following the advice of Kalliamvakou et al.~\cite{Kalliamvakou2014}, we further excluded repositories that were not used for software development by inspecting the programming languages automatically detected for each repository by GitHub. If no programming language was detected for a repository, we excluded this repository from our sample. 

We manually categorized the README files contained in our samples as end-user applications, frameworks, libraries, learning resources, and projects related to UI. The majority of our README files were related to end-user applications (i.e., 42\%) which includes client/server applications, apps/games, plugins, engines, databases, extensions, etc. The second largest category of files was related to libraries (27.9\%). Our sample also contained README files related to programming learning resources (17.4\%) such as tutorials, assignments, and labs. The remaining files were categorized as frameworks (7.3\%) and user interfaces (5.4\%) such as CSS styles and images. 

We also excluded repositories for which the README file was very small. We considered a file to be very small if it contained less than two kilobytes of data. This threshold was set based on manual inspection of the files which revealed that files with less than two kilobytes of content typically only contained the repository name, which is the default content of a new README file on GitHub.

During the manual annotation (see Section~\ref{sec:annotation}), we further excluded README files if their primary language was not English. Table~\ref{tab:readme-files} shows the number of repositories excluded based on these filters. Our final sample contains 393 README files, which results in a confidence interval of 4.94 at a confidence level of 95\% for our conclusions regarding the distribution of section types in the population of all GitHub repositories, assuming a population of 20 million repositories.

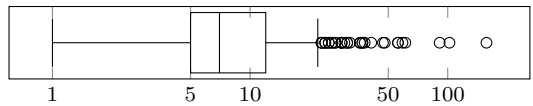
\begin{figure}
\centering
\begin{tikzpicture}
\begin{axis}[y=1cm, try min ticks=2, ytick=5, xmode=log, xtick={1,5,10,50,100}, xticklabels={1,5,10,50,100}]
\addplot[black, mark=o, boxplot, color=black]
table[row sep=\\,y index=0] {
data\\
26\\ 9\\ 37\\ 12\\ 6\\ 10\\ 9\\ 15\\ 7\\ 5\\ 12\\ 47\\ 9\\ 3\\ 13\\ 12\\ 8\\ 5\\ 38\\ 9\\ 17\\ 4\\ 12\\ 7\\ 12\\ 3\\ 4\\ 91\\ 23\\ 3\\ 4\\ 16\\ 12\\ 29\\ 6\\ 8\\ 11\\ 7\\ 7\\ 7\\ 6\\ 6\\ 6\\ 22\\ 7\\ 6\\ 6\\ 4\\ 18\\ 5\\ 7\\ 6\\ 6\\ 5\\ 7\\ 20\\ 6\\ 2\\ 9\\ 5\\ 10\\ 32\\ 4\\ 4\\ 1\\ 1\\ 9\\ 20\\ 11\\ 1\\ 19\\ 13\\ 16\\ 14\\ 3\\ 13\\ 7\\ 3\\ 5\\ 14\\ 6\\ 15\\ 7\\ 5\\ 8\\ 5\\ 8\\ 7\\ 13\\ 5\\ 5\\ 9\\ 20\\ 7\\ 7\\ 6\\ 23\\ 9\\ 8\\ 16\\ 1\\ 5\\ 27\\ 15\\ 1\\ 12\\ 1\\ 9\\ 5\\ 59\\ 1\\ 7\\ 14\\ 37\\ 7\\ 9\\ 3\\ 10\\ 8\\ 8\\ 22\\ 7\\ 10\\ 5\\ 23\\ 10\\ 6\\ 18\\ 6\\ 4\\ 10\\ 4\\ 11\\ 4\\ 19\\ 13\\ 10\\ 12\\ 16\\ 11\\ 9\\ 7\\ 4\\ 29\\ 4\\ 9\\ 19\\ 18\\ 13\\ 11\\ 7\\ 6\\ 6\\ 9\\ 9\\ 4\\ 1\\ 9\\ 3\\ 6\\ 3\\ 12\\ 3\\ 13\\ 7\\ 12\\ 4\\ 10\\ 7\\ 15\\ 6\\ 9\\ 14\\ 2\\ 7\\ 1\\ 7\\ 10\\ 14\\ 157\\ 56\\ 1\\ 6\\ 5\\ 6\\ 8\\ 5\\ 3\\ 4\\ 5\\ 4\\ 10\\ 4\\ 13\\ 6\\ 3\\ 5\\ 6\\ 9\\ 6\\ 5\\ 6\\ 14\\ 2\\ 9\\ 5\\ 3\\ 11\\ 5\\ 8\\ 3\\ 3\\ 7\\ 12\\ 16\\ 6\\ 1\\ 4\\ 6\\ 19\\ 6\\ 2\\ 11\\ 13\\ 12\\ 2\\ 24\\ 11\\ 7\\ 3\\ 9\\ 8\\ 6\\ 4\\ 5\\ 9\\ 5\\ 19\\ 5\\ 7\\ 6\\ 10\\ 10\\ 3\\ 29\\ 12\\ 3\\ 11\\ 8\\ 61\\ 31\\ 1\\ 10\\ 19\\ 41\\ 4\\ 5\\ 24\\ 7\\ 17\\ 6\\ 102\\ 6\\ 4\\ 8\\ 8\\ 5\\ 6\\ 3\\ 4\\ 8\\ 25\\ 12\\ 5\\ 3\\ 5\\ 14\\ 8\\ 2\\ 4\\ 18\\ 7\\ 5\\ 2\\ 5\\ 1\\ 18\\ 6\\ 15\\ 4\\ 19\\ 31\\ 12\\ 6\\ 3\\ 2\\ 2\\ 10\\ 3\\ 6\\ 13\\ 1\\ 2\\ 15\\ 15\\ 4\\ 8\\ 6\\ 27\\ 4\\ 7\\ 10\\ 7\\ 9\\ 9\\ 13\\ 4\\ 4\\ 6\\ 20\\ 10\\ 25\\ 6\\ 4\\ 5\\ 3\\ 3\\ 9\\ 9\\ 8\\ 48\\ 4\\ 56\\ 36\\ 14\\ 3\\ 26\\ 11\\ 18\\ 30\\ 6\\ 7\\ 29\\ 3\\ 8\\ 5\\ 24\\ 4\\ 7\\ 2\\ 7\\ 5\\ 3\\ 1\\ 8\\ 18\\ 9\\ 6\\ 7\\ 36\\ 1\\ 3\\ 9\\ 3\\ 20\\ 9\\ 5\\ 14\\ 7\\ 5\\ 1\\ 14\\ 7\\ 4\\ 1\\ 2\\ 6\\ 15\\ 9\\ 6\\ 5\\ 2\\ 8\\ 7\\ 2\\ 10\\ 8\\ 11\\ 10\\ 8\\ 7\\ 4\\ 14\\ 
};
\end{axis}
\end{tikzpicture}
\caption{Number of sections per README file in our sample}
\label{fig:sections-per-file}
\end{figure}

We then used GitHub's markdown\footnote{\url{https://guides.github.com/features/mastering-markdown/}} to extract all sections from the README files in our sample, yielding a total of 4,226 sections distributed over the 393 README files. GitHub's markdown offers headers at different levels (equivalent to HTML's \texttt{h1} to \texttt{h6} tags) for repository owners to structure their README files. Figure~\ref{fig:sections-per-file} shows the distribution of the number of sections per README file. The median value is seven and 50\% of the files contain between five and twelve sections.

\subsection{Coding schema}

We adopted `open coding' since it is a commonly used methodology to identify, describe, or categorize phenomena found in qualitative data~\cite{Corbin1990}. In order to develop a coding scheme, one author manually classified a random sample of fifty README files into meaningful categories (known as codes~\cite{Miles1994}). Our findings from this examination consist of a tentative list of seven categories (e.g. what, why, how) and sub categories (e.g., introduction, background). After defining initial codes, we trialed them on 150 README sections using two annotators. For this round of coding, we obtained inter-rater reliability of 76\%. Following this trial, we refined our codes until we reached agreement on a scheme that contained codes for all of the types of README sections we encountered. Finally, we define the `other' category only when all other possibilities have been exhausted. Table~\ref{tab:readme-reference} shows the finalized set of categories as well as example section headings for each category found in this initial sample of README files. The categories roughly correspond to the content of README files that is recommended by GitHub (cf. Introduction). 

We identified the first category (`What') based on headings such as `Introduction' and `About', or based on the text at the beginning of many README files. We found that either a brief introduction or a detailed introduction is common in our dataset. Conversely, category two (`Why') is rare in README files. For instance, some repositories compare their work to other repositories based on factors such as simplicity, flexibility, and performance. Others list advantages of their project in the introduction. 

The most frequent category is `How' since the majority of README files tend to include instructions on \textit{how to use the project} such as programming-related content (e.g., configuration, installation, dependencies, and errors/bugs). Table~\ref{tab:readme-reference} lists a sample of section headings that belongs to the `How' category. Further, it is also important to the reader of a README file to be familiar with the status of the project, including versions as well as complete and in-progress functionality. We categorize this kind of time-related information into the fourth code (`When'). 

We categorize sections as `Who' content when they include information about who the project gives credit to. This could be the project team or acknowledgements of other projects that are being reused. This category also includes information about licence, contact details, and code of conduct. The second most frequent category is `References'. This category includes links to further details such as API documentation, getting support, and translations. This category also includes `related projects', which is different from the `comparison with related projects' in category `Why' due to the lack of an explicit comparison. Our final category is `Contribution', which includes information about how to fork or clone the repository, as well as details on how to contribute to the project. Our manual analysis indicated that some repositories include separate \texttt{CONTRIBUTING.md} files which contain instructions on how to get involved with the project. We do not consider \texttt{CONTRIBUTING.md} files in this study. In addition, we included a category called `Other' which is used for sections that do not belong to any of the aforementioned seven categories.

\begin{table}
\centering
\caption{README section coding reference}
\label{tab:readme-reference}
\begin{tabular}{rll}
\toprule
\# & Category & Example section headings \\
\midrule
1 &	What &	Introduction, project background  \\
\midrule
2 &	Why &	\makecell[l]{Advantages of the project, \\ comparison with related work} \\
\midrule
3 &	How &	\makecell[l]{Getting started/quick start,  how to run, \\ installation, how to update, configuration, \\setup, requirements, dependencies, \\ languages, platforms, demo, \\ downloads, errors and bugs} \\
\midrule
4 &	When &	\makecell[l]{Project status, versions, \\ project plans, roadmap}  \\
\midrule
5 &	Who	& \makecell[l]{Project team, community, \\ mailing list, contact, acknowledgement, \\licence, code of conduct } \\
\midrule
6 &	References & \makecell[l]{API documentation, getting support,\\feedback, more information, \\translations, related projects} \\
\midrule
7 &	Contribution &	Contributing guidelines  \\
\midrule
8 &	Other &  \\
\bottomrule
\end{tabular}
\end{table}

\subsection{Manual annotation}
\label{sec:annotation}

We initially used two annotators to code the dataset. One of the annotators was a PhD candidate specializing in Software Engineering while the other one is an experienced Software Engineer working in industry. Neither of the annotators is an author of this paper. Each annotator spent approximately thirty hours to annotate the dataset. The task of an annotator is to read the section headings and contents and assign a code based on the coding reference. The annotators assign codes from the eight available codes (Table~\ref{tab:readme-reference}). Each section of a README file can have one or more codes.

We measured the inter-rater agreement (i.e. Kappa) between the two annotators and obtained an agreement of 0.858. We used a third annotator to rectify the sections which had no agreement. For this, two authors of the paper (Software Engineering academics) co-annotated the remaining sections that had no agreement. For all cases, we then used a majority vote to determine the final set of codes for each section, i.e., all codes that had been used by at least two annotators for a section were added to the final set of codes for that section.\footnote{In cases where there was perfect agreement between the two annotators, the majority vote rule simply yields the codes that both annotators agreed on.} In very few cases, there was still no agreement on any set of codes after considering the codes from three annotators. These cases were manually resolved by discussion between two authors of this paper.

We manually examined the instances where the annotators disagree. Annotators were likely confused when the README file includes `Table Of Contents (TOC)' as they have provided inconsistent codes in these instances. Since TOC is included at the beginning of the file, one annotator considers it as category `What' while the other one placed it in the references. However, the third annotator categorized TOC into `Other', which is what we used in the final version of the annotated dataset.  Another common confusion occurred when categorizing `community-related' content. Our coding reference (Table~\ref{tab:readme-reference}) suggests that community-related information should be placed in the `Who' category. However, one annotator identified it in the `Contribution' category. We generally resolved `community-related' disagreements by placing them into the `Who' category, in accordance with our coding guide. 

We also noticed that our annotators are reluctant to place content into the `Other' category. Instead, they attempted to classify README contents into the other seven categories. Further, one of the main reasons for disagreement was the inclusion of external links as section titles or contents. For example, one README file listed the middleware available to use with their project as section titles. However, these section titles include ``Apache'' and ``Nginx''.\footnote{\url{https://github.com/microlv/prerender}} One annotator categorized these sections into `How' while the other placed them in additional resources (code `References') since they have external links. There can be multiple headings which depend on this decision. For instance, one README file contained 36 headings about configurations. They are categorized into `How' by one annotator while the other one placed them in additional resources since they have URLs. Resolving this disagreement affected many sections at once. 

Further, some README files include screenshots or diagrams to provide an overview or demonstrations. These are expected to be classified in `Other'. However, annotators have occasionally assigned codes such as `What', `How', and `References' to image contents. Another challenging decision occurs when repositories include all the content under a single heading. This causes the annotators to assign multiple codes which possibly do not overlap between annotators. In addition, we sometimes found misleading headings such as `how to contribute' where the heading would suggest that the content belongs to category `Contribution'. However, in a few cases, the content of this section included information on `how to use the project' (i.e., download, install, and build). 

\section{The content of GitHub README files}\label{sec:readme-content}

Table~\ref{tab:distribution-of-codes} demonstrates the distribution of categories based on the human annotation (column 3 on `sections') and the README files in our sample (column 4 on `files'). Based on manually annotated sections, the most frequent category is `How' (58.4\%), while the least frequent was `Other' (1.4\%).  As mentioned previously, as part of the coding, our annotators also excluded non-English content that had not been detected by our automated filters (code `-'). The same applies to parts of README files that had been incorrectly detected as sections by our automated tooling.

\begin{table}
\centering
\caption{Distribution of README categories; App: end-user applications; Lib: libraries; Frame: frameworks; Learn: learning resources, UI: user interfaces}
\label{tab:distribution-of-codes}
\begin{tabular}{llrrrrrrr}
\toprule
\# & Category & \# Sections & \# Files & App & Lib & Frame & Learn & UI \\
& & (\%) & (\%) & (\%) & (\%) & (\%) & (\%) & (\%) \\
\midrule
1 &	What & 707  & 381 & 14.0 & 14.2 & 12.3 & 22.6 & 9.6 \\
& & (16.7\%)  & (97.0\%) \\
2 &	Why & 116 & 101 & 2.6 & 2.4 & 3.2 & 2.6 & 0.3 \\
& & (2.7\%) & (25.7\%) \\
3 &	How & 2,467  & 348 & 49.5 & 45.0 & 52.9 & 52.9 & 65.6  \\
& & (58.4\%) & (88.5\%) \\
4 &	When &	180  & 84 & 5.8 & 2.5 & 4.4 & 0.6 & 1.3  \\
& & (4.3\%) & (21.4\%) \\
5 &	Who & 322  & 208 & 6.6 & 9.5 & 5.9 & 3.7 & 6.3  \\
& & (7.6\%) & (52.9\%) \\
6 &	References &	858  & 239 & 18.4 & 22.2 & 17.2 & 13.5 & 10.3  \\
& & (20.3\%) & (60.8\%) \\
7 &	Contribution &	122  & 109 & 2.4 & 2.7 & 3.2 & 1.6 & 2.6  \\
& & (2.9\%) & (27.8\%) \\
8 &	Other &	58  & 27 & 0.5 & 1.4 & 0.7 & 2.3 & 3.9  \\
& & (1.4\%) & (6.9\%) \\
\midrule
- &	Exclusion &	696 &	 \\
\bottomrule
\end{tabular}
\end{table}

\begin{table}
\centering
\caption{Quantity of codes per section}
\label{tab:multiple-codes}
\begin{tabular}{rr}
\toprule
\# Codes & \# Sections \\
\midrule
5 &	2 \\
4 &	6 \\
3 &	40 \\
2 &	498 \\
1 &	3,680 \\
\midrule
Total & 4,226 \\
\bottomrule
\end{tabular}
\end{table}


Based on the consideration of files in our sample (fourth column of Table~\ref{tab:distribution-of-codes}), 97\% of the files contain at least one section describing the `What' of the repository and 88.5\% offer some `How' content. Other categories, such as `Contribution', `Why', and `Who', are much less common.

The last five columns of Table~\ref{tab:distribution-of-codes} demonstrate the distribution of codes across various file types (e.g., end-user applications, libraries). The most common code among all file types is `How' while `What' and `References' are common in all file types except README files related to `user interfaces'. Further, learning related resources such as assignments and tutorials rarely contain information related to `When' and `Contribution'.

Further, we report the distribution of number of codes across the sections of GitHub README files in our sample (Table~\ref{tab:multiple-codes}). The sections that are annotated using four or five codes mostly stem from README files that only contain a single section. Interestingly, the majority of these files include `What', `Who', and `References'.  Also, 92\% of the sections which are annotated using three codes include `What'.  Unsurprisingly, the most popular combination of two codes was `How' and `References', enabling access to additional information when learning `how to use the project'. These relationships are further explored in the following section. 

\subsection{Relations between codes}

As with any qualitative coding schema, there may be some overlap between the different types of sections outlined in our coding reference (cf.~Table \ref{tab:readme-reference}). For example, API documentation, which the coding reference shows as an example for `References' is often also related to `How' or could be related to `Contribution'. To systematically investigate the overlap between different section types based on the manually annotated data, we applied association rule learning~\cite{AR} to our data using the \texttt{arules} package in R. To find interesting rules, we grouped the data both by sections (i.e., each section is a transaction) and by files (i.e., each file is a transaction).

\begin{table}
\centering
\caption{Association rules at section level}
\label{tab:section-rules}
\begin{tabular}{lllrr}
\toprule
Rule & & & Support & Confidence \\
\midrule
\{Why, How\} & $\Rightarrow$ & \{What\} & 0.002 & 1.00 \\
\{Why, References\} & $\Rightarrow$ & \{What\} & 0.003 & 0.93 \\
\bottomrule
\end{tabular}
\end{table}

\begin{table}
\centering
\caption{Association rules at file level}
\label{tab:file-rules}
\begin{tabular}{lllrr}
\toprule
Rule & & & Support & Confidence \\
\midrule
\{Who\} & $\Rightarrow$ & \{What\} & 0.52 & 0.98 \\
\{How, References\} & $\Rightarrow$ & \{What\} & 0.54 & 0.98 \\
\{References\} & $\Rightarrow$ & \{What\} & 0.59 & 0.97 \\
\{How\} & $\Rightarrow$ & \{What\} & 0.86 & 0.97 \\
\{References\} & $\Rightarrow$ & \{How\} & 0.55 & 0.91 \\
\{What, References\} & $\Rightarrow$ & \{How\} & 0.54 & 0.91 \\
\{What\} & $\Rightarrow$ & \{How\} & 0.86 & 0.89 \\
\bottomrule
\end{tabular}
\end{table}

Table~\ref{tab:section-rules} shows the extracted rules at section level. We only consider rules with a support of at least 0.0013 (i.e., the rule must apply to at least five sections) and a confidence of at least 0.8. Due to the small number of sections for which we assigned more than one code, only two rules were extracted: Sections that discuss the `Why' and `How' are likely to also contain information on the `What'. Similarly, sections that discuss the `Why' of a project and contain `References' are also likely to contain information on the `What'.

At file level, we were able to find more rules, see Table~\ref{tab:file-rules}. For these rules, we used a minimum support of 0.5 and a minimum confidence of 0.8. We chose a minimum support of 0.5 to limit the number of rules to the most prevalent ones which are supported at least by half of the README files in our dataset. The rules extracted with these parameters all imply `What' or `How' content to be present in a README file. For example, we have a 98\% confidence that a file that contains information about `Who' also contains information about the `What' of a project. This rule is supported by 52\% of the README files in our dataset.

\subsection{Examples}

In this section, we present an example for each of the categories to illustrate the different codes.

\paragraph{What.} The leading section of the GitHub README file of the \texttt{ParallelGit} repository\footnote{\url{https://github.com/jmilleralpine/ParallelGit}} by GitHub user \texttt{jmilleralpine} is a simple example of a section that we would categorize into the `What' category. The section header simply restates the project name (``ParallelGit'') and is followed by this brief description: ``A high performance Java JDK 7 nio in-memory filesystem for Git.'' Since this is an introduction to the project, we assign the code `What'.

\paragraph{Why.} The README file of the same repository (\texttt{ParallelGit}) also contains a section with the heading ``Project purpose explained'' which we categorize into the `Why' category. This section starts with a list of four bullet points outlining useful features of Git, followed by a brief discussion of the ``lack of high level API to efficiently communicate with a Git repository''. The README file then goes on to explain that ``ParallelGit is a layer between application logic and Git. It abstracts away Git's low level object manipulation details and provides a friendly interface which extends the Java 7 NIO filesystem API.'' Since this section describes the purpose of the project and motivates the need for it, we assign the code `Why'.

\paragraph{How.} The same README file also contains a section with the heading ``Basic usages'', which we classify into the `How' category. It provides two short code snippets of seven and eight lines, respectively, which illustrate the use cases of ``Copy a file from repository to hard drive'' and ``Copy a file to repository and commit''. We assign the code `How' because this section explains how to run the software.

\paragraph{When.} An example of a section discussing the `When' aspect of a project is given by the section with the heading ``Caveats'' of the \texttt{Sandstorm} repository\footnote{\url{https://github.com/solomance/sandstorm}} by GitHub user \texttt{solomance}. The project is a self-hostable web app platform. In its ``Caveats'' section, the README file states ``Sandstorm is in early beta. Lots of features are not done yet, and more review needs to be done before relying on it for mission-critical tasks. That said, we use it ourselves to get work done every day, and we hope you'll find it useful as well.'' Since this section describes the project status, we assign the code `When'.

\paragraph{Who.} Going back to the README file of the \texttt{ParallelGit} repository, it concludes with a section with the heading ``License'' and the following text: ``This project is licensed under Apache License, Version 2.0.'' A link to the license text is also included. We categorized this section under `Who' since it contains licence information (see Table~\ref{tab:readme-reference}).

\paragraph{References.} The previously mentioned README file of the \texttt{Sandstorm} repository also contains sections that we categorized as `References', e.g., the section with the heading ``Using Sandstorm''. This section only contains the statement ``See the overview in the Sandstorm documentation'' which links to more comprehensive documentation hosted on \url{https://docs.sandstorm.io/}. We assign the code `References' since the section does not contain any useful content apart from the link to more information. This section showcases one of the challenges of classifying the content of sections contained in GitHub README files: While the section header suggests that the section contains `How' information, the body of the section reveals that it simply contains a link.

\paragraph{Contribution.} The README file of \texttt{Sandstorm} also contains a section with the heading ``Contribute'' which we categorized under `Contribution'. The section states ``Want to help? See our community page or get on our discussion group and let us know!'' and contains links to a community page hosted on \url{https://sandstorm.io/} as well as a discussion group hosted on Google Groups.\footnote{\url{https://groups.google.com/}} We assign the code `Contribution' rather than `References' since this section contains information other than links, i.e., the different ways in which contributions can be made. Arguably, this is a corner case in which the code `References' would also be justifiable.

\paragraph{Other.} An example of a section that we were not able to categorize using any of the previous seven categories is the last section in the README file of the \texttt{Blackjack} repository\footnote{\url{https://github.com/ChadLactaoen/Blackjack}} by GitHub user \texttt{ChadLactaoen}. The section does not contain any content and simply consists of the section heading ``Have fun!'' In this case, the section feature of GitHub markdown was used for highlighting rather than for structuring the content of the README file. We therefore categorized the section as `Other'.

\begin{tcolorbox}
RQ1: Section content of GitHub README files can be categorized into eight types, with the `What' and `How' content types being very common and information on project status being rare.
\end{tcolorbox}

\section{A GitHub README Content Classifier}\label{sec:classifier}
In this section, we describe our automated classification approach for classifying GitHub README content. We first describe the overall framework of our approach and then explain each of its steps. For the development of this classifier, we use the set of sections associated with one of 8 classes along with sections labeled `Exclusion', and split the dataset into two, a {\em development} set comprising 25\% of the data, and an {\em evaluation} set comprising 75\% of the data. We analyze and use the {\em development} set to design features for the classifier, such as heuristics based on language patterns (see Section \ref{sec:heuristic_features}). The {\em evaluation} set is the hold out set that is used for evaluation of the classifier through ten-fold cross-validation. A similar process of dividing a dataset into two -- one for manual analysis for feature identification, and another for evaluation -- has been done in prior studies (e.g., \cite{panichella2015can}) to improve reliability of reported results. Our code, dataset, along with scripts for the experiments as well as a README file containing information
on how to use them are available at \url{https://github.com/gprana/READMEClassifier}

\subsection{Overall Framework}
\begin{figure}[hbt]
\centering
\vspace{-1cm}
\includegraphics[width = 0.9\textwidth]{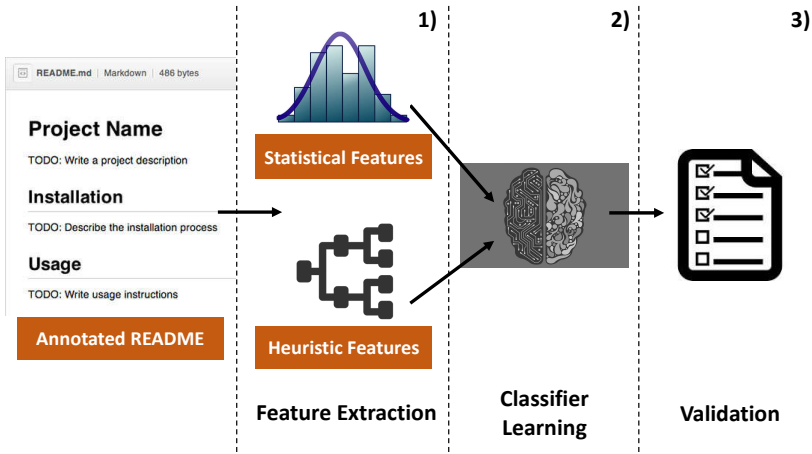}
\vspace{-1cm}
\caption{The overall framework of our automated GitHub README content classifier. }
\label{fig:framework}
\end{figure}
We present the overall framework of our automated classification approach in Figure~\ref{fig:framework}. The framework consists of the following steps:
\begin{enumerate}
    \item {\bf Feature Extraction:} From each section of the annotated GitHub README files, we extract meaningful features that can identify categories of a section's content. We extract statistical and heuristic features. These features are output to the next step for learning.
    \item {\bf Classifier Learning:} Using features from the previous step, we learn a classifier that can identify the categories that the content of each section belongs to. Since each section can belong to many categories, we use a multi-label classifier, which can output several categories for each section.
    \item {\bf Validation:} To choose our classifier setting, we need to validate our classifier performance on a hold out set. We experiment with different settings and pick the classifier that performs the best on the hold out set.
\end{enumerate}
We explain details of the above steps in the next subsections.

\subsection{Feature Extraction}
From the content of each section, we extract two sets of features: statistical features and heuristic features.
\subsubsection{Statistical Features}
These features compute word statistics of a README section. These features are constructed from combination of both heading and content of the section. To construct these features, the section's content and heading are first preprocessed. We perform two preprocessings:  {\em content abstraction} and {\em tokenization}. Content abstraction abstracts contents to their types. We abstract the following types of section content: {\em mailto} link, hyperlink, code block, image, and numbers. Each type is abstracted into a different string ({\em @abstr\_mailto}, {\em @abstr\_hyperlink}, {\em @abstr\_code\_section}, {\em @abstr\_image} and {\em @abstr\_number}, respectively). Such abstraction is performed since for classification, we are more interested in existence of those types in a section than its actual content. For example, existence of a source code block in a section may indicate that the section demonstrates usage of the project, regardless of the source code. With abstraction, all source code blocks are converted to the same string, and subsequently, into the same statistical feature. This abstraction is followed by tokenization, which converts a section into its constituent words, and English stop word removal. For the stop word removal, we use the stop words provided by {\em scikit-learn} \cite{pedregosa2011scikit}. 

After preprocessing, we count the number of times a word appears in each section. This is called the Term Frequency (TF) of a word in a section. If there are $n$ words that appear in the set of sections used for training the classifier (after preprocessing), we would have $n$ statistical features for each section. If a word does not appear in a section, then its TF is zero. We also compute the Inverse Document Frequency (IDF) of a word. IDF of a word is defined as the reciprocal of the number of sections in which the word appears. We use a multiplication of TF and IDF as an information retrieval feature for a particular word.

\subsubsection{Heuristic Features}
\label{sec:heuristic_features}
There has been work such as Panichella et al.~\cite{panichella2015can} which exploits recurrent linguistic patterns within a category of sentences to derive heuristics that can aid classification. Given this, we manually inspected the content of various sections in the {\em development} set to try to identify patterns that may be useful to distinguish each category. The following are the resulting heuristic features that we use for the classifier.
\begin{enumerate}
    \item {\bf Linguistic Patterns:} This is a binary feature that indicates whether a particular linguistic pattern exists in a section. We discover linguistic patterns by looking at words/phrases that either appear significantly more in one particular category or are relatively unique to a particular category. A linguistic pattern is tied to either a section's heading or content. A pattern for heading is matched only to the section's heading. Similarly, a pattern for content is matched only to the section's content. There are 55 linguistic patterns that we identified.\footnote{The linguistic patterns are available in \url{https://github.com/gprana/READMEClassifier/blob/master/doc/Patterns.ods}.} 
    \item {\bf Single-Word Non-English Heading:} This is a binary feature that indicates whether a section's heading is a single word non-English heading. An example is a method name, which may be used as heading in a section describing the method and usually belongs to the `How' category. This check is performed by checking the word against the wordlist corpora from NLTK~\cite{bird2009natural}.
    \item {\bf Repository Name:} This is a binary feature that indicates whether any word in the repository name is used in a section's heading. This is based on the observation that  the README section that provides an overview of the project likely contains common words from the project name. For example, a repository of a project called `X' will contain `X' in its name, and the README section providing an overview of the project may be given a heading along the lines of `About X', `Overview of X', or `Why X'. This is different, for example, from README sections containing licence information or additional resources. 
    \item {\bf Non-ASCII Content Text:} This is a binary feature that indicates whether a section contains any non-ASCII character. It is based on the observation that README sections containing text written in non-ASCII characters tend to be categorized as `Exclusion', although they often also contain parts (e.g., technical terms or numbers) written in ASCII characters.
\end{enumerate}

\subsection{Classifier Learning}
Given the set of features from the previous step, we construct a multi-label classifier that can automatically categorize new README sections. We use a binary relevance method for multi-label classification~\cite{luaces2012binary}. This method transforms the problem of multi-label classification into a set of binary classifications, with each binary classification performed for one label independently from the other labels. Due to the small number of entries in the `Why' category, combined with the fact that a large proportion of content in this category is also assigned to the `What' category, we combined the two categories. We therefore ended up with eight categories including `Exclusion', and subsequently created eight binary classifiers, each for a particular category. 

A binary classifier for a particular label considers an instance that contains the label as a positive instance, otherwise it is a negative instance. As such, the training set for the binary classifier is often imbalanced. Thus, we balance the training set by performing oversampling. In this oversampling, we duplicate instances of minority classes and make sure that each instance is duplicated roughly until we have the same number of positive and negative instances in the set.

\subsection{Validation}
In this step, we determine the classifier setting by performing ten-fold cross validation. The setting that leads to the highest classifier performance is selected as final setting. 


\section{Evaluation of the Classifier}
\label{sec:evaluation}
We conduct experiments with our SVM-based classifier on the dataset annotated in Section~\ref{sec:readme-content}. We evaluate the classifier on the {\em evaluation} set using ten-fold cross validation. We follow our framework in Section~\ref{sec:classifier} to construct our classifier. For evaluation, the TF-IDF vocabulary is constructed from the {\em evaluation} set, and is not shared with the {\em development} set. The size of this vocabulary created from the {\em evaluation} set is 14,248. We experiment with the following classification algorithms: Support Vector Machine (SVM), Random Forest (RF), Logistic Regression (LR), Naive Bayes (NB), and $k$-Nearest Neighbors (kNN). We use implementations of the classification algorithms from {\em scikit-learn}~\cite{pedregosa2011scikit}. To evaluate the usefulness of the classification, we used the automatically determined classes to label sections in GitHub README files using badges and showed files with and without these badges to twenty software professionals. 

\subsection{Evaluation metric}

We measure the classification performance in terms of F1 score. 
F1 score for multi-label classification is defined below.
\[
    F1 = \frac{\sum_{l \in L}{w_l \times F1_l}}{|L|}
\]
\[
    F1_l = \frac{2 \times Precision_l \times Recall_l}{Precision_l+Recall_l}
\]
\noindent where $w_l$ is the proportion of the actual label $l$ in all predicted data. $F1_l$ is the F1 score for label $l$, $L$ is the set of labels, $Precision_l$ is precision for label $l$, and $Recall_l$ is the recall for label $l$. When computing precision/recall for label $l$, an instance having label $l$ is considered as a positive instance, otherwise it is a negative instance. Precision is the proportion of predicted positive instances that are actually positive while recall is the proportion of actual positive instances that are predicted as positive.

For this work we consider both precision and recall as equally important. Taking into account that each section can have a different mix of content, our goal is to maximize completeness of the label set assigned to a section while avoiding clutter that can result from assigning less relevant labels.

\subsection{Evaluation results}

The results of our evaluation are shown in Table~\ref{tab:validation}. Our experimental results show that our SVM-based classifier can achieve an F1 score of 0.746 on the {\em evaluation} set using ten-fold cross validation. We also experiment with using SMOTE~\cite{chawla2002smote} on the best performing (SVM-based) classifier to compare its effectiveness with the oversampling approach, and found that it resulted in a lower F1 of 0.738.

The per category F1 obtained from the SVM-based classifier is shown in Table~\ref{tab:f1}.

\begin{table}
\centering
\caption{Results for Different Classifiers}
\label{tab:validation}
\begin{tabular}{lc}
\toprule
Classifier & $F1$\\
\midrule
SVM & 0.746 \\ 
RF & 0.696 \\
NB & 0.518 \\
LR & 0.739 \\
kNN & 0.588 \\
\bottomrule
\end{tabular}
\end{table}

\begin{table}
\centering
\caption{Effectiveness of Our SVM-based Classifier}
\label{tab:f1}
\begin{tabular}{lccc}
\toprule
Category & F1 & Precision & Recall\\
\midrule
What and Why & 0.615 & 0.627 & 0.604\\
How & 0.861 & 0.849 & 0.874\\
When & 0.676 & 0.669 & 0.683\\
Who & 0.758 & 0.810 & 0.711\\
References & 0.605 & 0.606 & 0.603\\
Contribution & 0.814 & 0.857 & 0.774\\
Other & 0.303 & 0.212 & 0.537\\
Exclusion & 0.674 & 0.596 & 0.775\\
\midrule
Overall & 0.746 & 0.742 & 0.759\\

\bottomrule
\end{tabular}
\end{table}

In addition to F1, we measured the performance of our classification using Kappa~\cite{landis1977measurement}, ROC AUC~\cite{fawcett2006introduction}, and MCC~\cite{boughorbel2017optimal}. Our classifier can achieve a weighted average Kappa of 0.831, a weighted average ROC AUC of 0.957, and a weighted average MCC of 0.844. As prior work (e.g.,~\cite{lessmann2008benchmarking,prasetyo2012automatic,romano2011using,xia2014towards}) consider F-measure and/or AUC of  0.7 or higher to be reasonable, we believe the evaluation result demonstrates that the SVM-based classifier design has sufficiently good performance.

\begin{tcolorbox}
RQ2: We can automatically classify content of sections in GitHub README files with F1 of 0.746
\end{tcolorbox}

\subsection{Speed}
We evaluate the speed of the best performing SVM-based classifier using a test machine with the following specifications: Intel Core i7-4710HQ 2.50 GHz CPU, 16 GB RAM laptop with SSD storage and Windows 10 64-bit. For this part of the evaluation, input data comprise the combined set README files from development and evaluation sets. We find that training of the classifier on this combined set takes 181 seconds. Afterwards, the classifier is able to label sections in a given input README file in less than a second. This indicates that the classifier is fast enough for practical use.

\subsection{Multi-category sections vs.~single-category sections}

We expect that classifying multi-category sections is harder than classifying single-category sections. To confirm this, we exclude sections that belong to more than one category. We perform a similar experiment using ten-fold cross validation. Our experimental results show that our SVM-based classifier achieves an F1 score of 0.773, which confirms that classifying single-category sections is indeed easier, although not by a significant margin.

\subsection{Usefulness of statistical vs.~heuristic features}

To investigate the value of a set of features, we remove the set and observe the classifier performance after such removal. Table~\ref{tab:feature-contribution} shows the classifier performance when we remove different sets of features. We observe performance reduction when removing any set of features. Thus, all sets of features are valuable for classifying README sections. Among the sets of features, the statistical features are more important since their removal reduces F1 far more as compared to removing heuristic features.
\begin{table}
\centering
\caption{Contribution of Different Sets of Features}
\label{tab:feature-contribution}
\begin{tabular}{ll}
\toprule
Set of Features Used & F1\\
\midrule
Only Heuristic & 0.584 \\
Only Statistical & 0.706 \\
\bottomrule
\end{tabular}
\end{table}

\subsection{Usefulness of particular features}

We are also interested in identifying which particular feature is more useful when predicting different categories. Using an SVM classifier, usefulness of a feature can be estimated based on the weight that the classifier assigns to the feature. For each category in the testing data, we consider an instance belonging to the category as a positive instance, otherwise it is  a negative instance. We learn an SVM classifier to get the weight of each feature. To capture significantly important features, we perform the Scott-Knott ESD (Effect Size Difference) test~\cite{tantithamthavorn2017mvt}. For the purpose of this test, we perform ten times ten-fold cross validation where each cross validation generates different sets. Thus, for each category and feature pair, we have 100 weight samples. We average the weights and run Scott-Knott ESD test on the top-5 features' weights. We present the result for each category in Figure~\ref{fig:top-features}. Features grouped by the same color are considered to have a negligible difference and thus have the same importance. 

Based on the observation, heuristics based on sections' headings appear to be useful in predicting categories. For example, {\em heur\_h\_k\_012} (check whether a lower cased heading contains the string `objective') is the second most useful features for predicting the `What and Why' category, while {\em heur\_h\_k\_006} (check whether a lower cased heading contains the string `contrib') is the third most useful feature for predicting the `Contribution' category. For the `Who' category, {\em heur\_h\_k\_007} (check whether a lower cased heading contains `credit') is the fifth most useful feature for prediction. Abstraction also appears to be useful, with {\em @abstr\_number} being the fifth ranking feature for predicting the `When' category. A possible reason is that the `When' category covers version history, project plans, and project roadmap, which often contain version number, year, or other numbers.


\begin{figure*}%
\centering
\subfigure{%
\label{fig:what-and-why-features-rank}%
\includegraphics[width=0.5\textwidth]{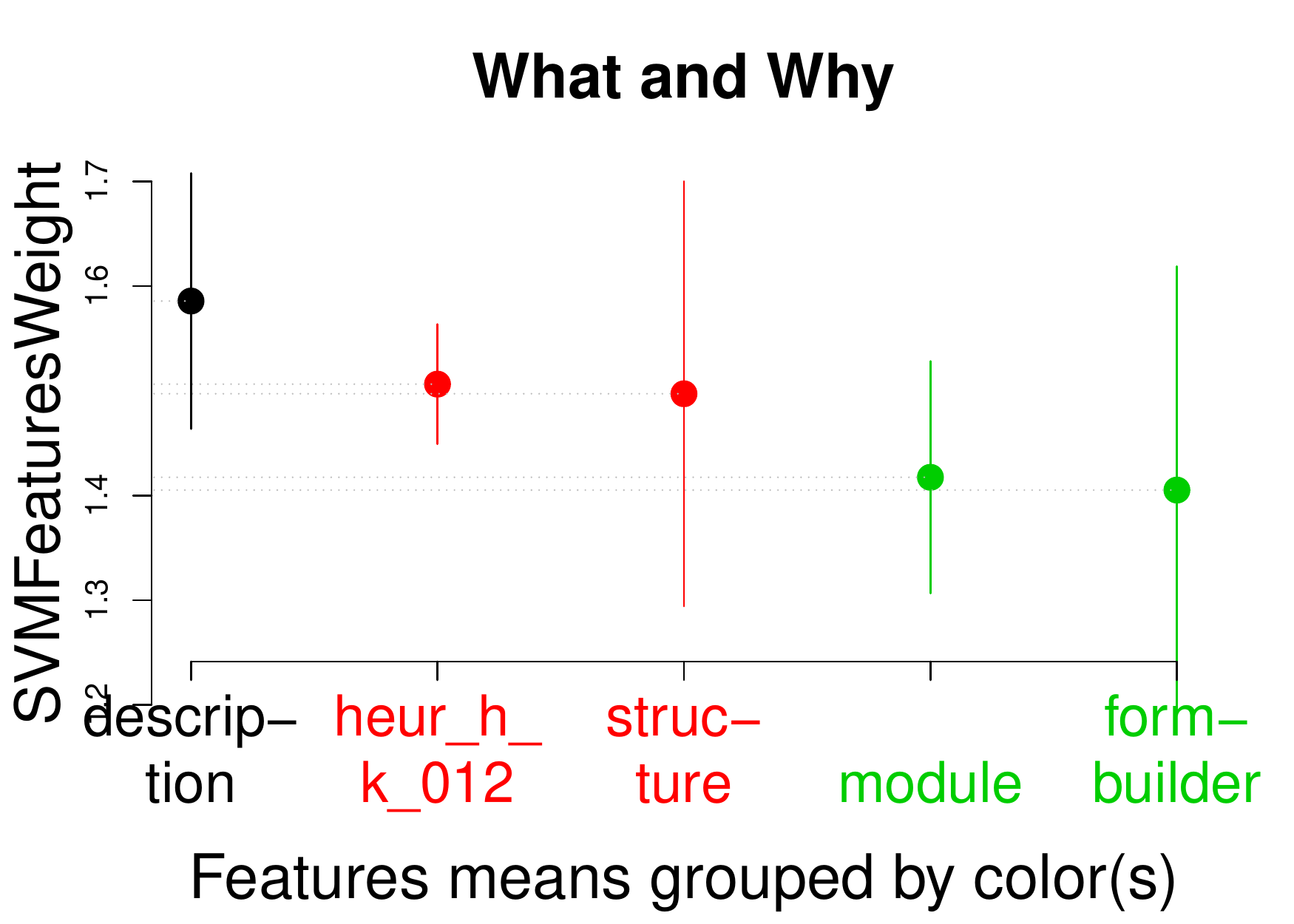}}%
\subfigure{%
\label{fig:how-features-rank}%
\includegraphics[width=0.5\textwidth]{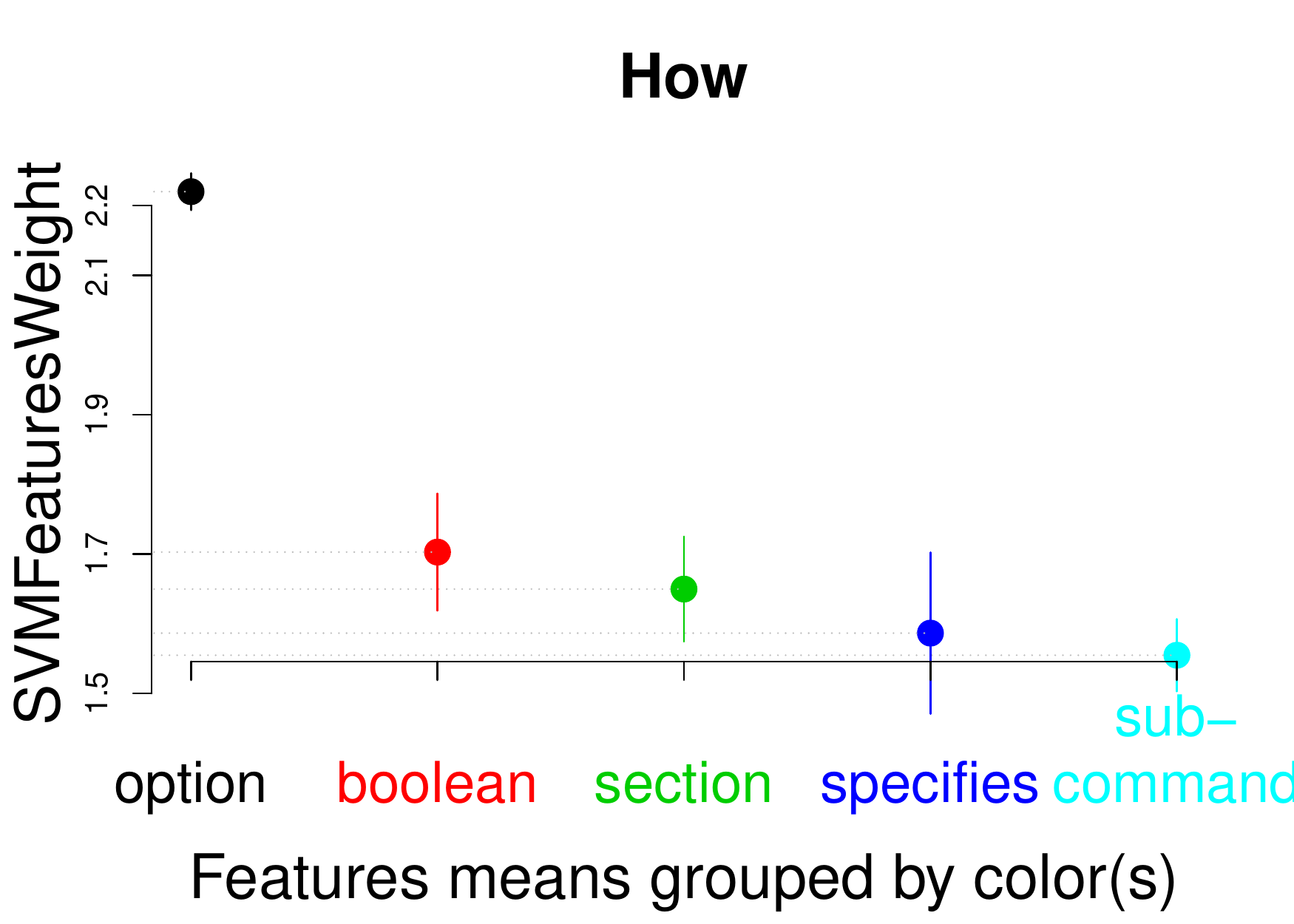}}%
\qquad
\subfigure{%
\label{fig:when-features-rank}%
\includegraphics[width=0.5\textwidth]{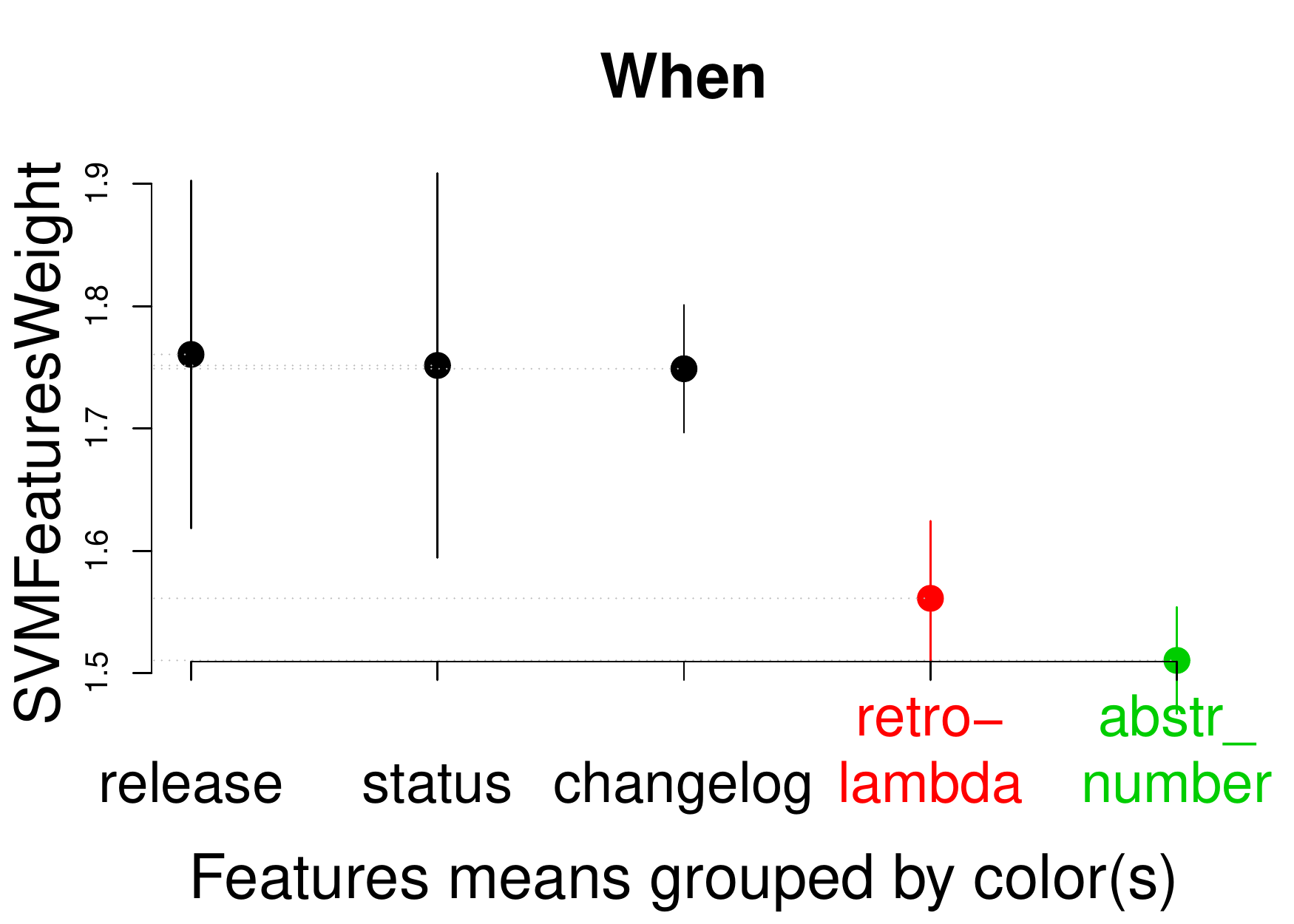}}%
\subfigure{%
\label{fig:who-features-rank}%
\includegraphics[width=0.5\textwidth]{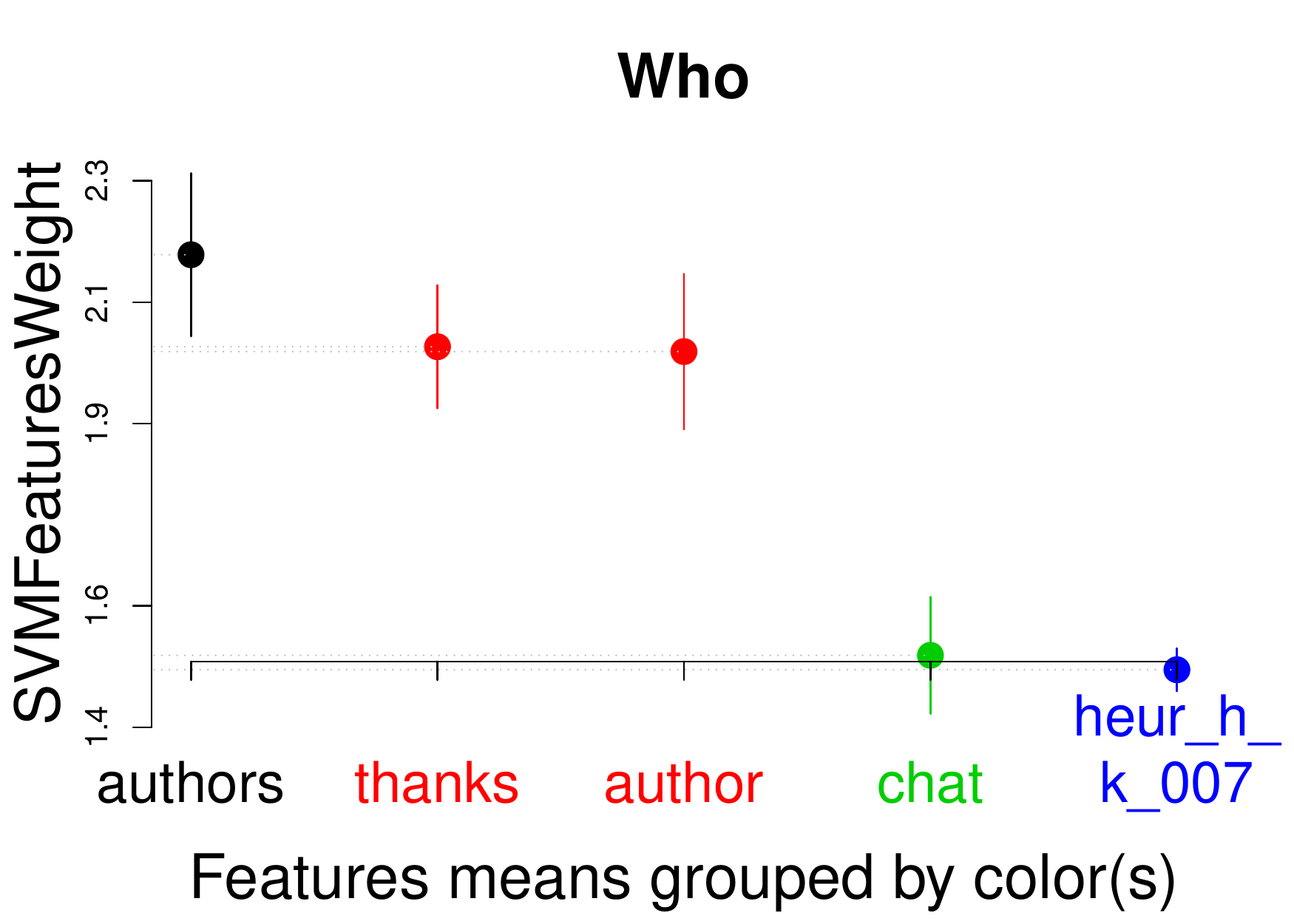}}%
\qquad
\subfigure{%
\label{fig:references-features-rank}%
\includegraphics[width=0.5\textwidth]{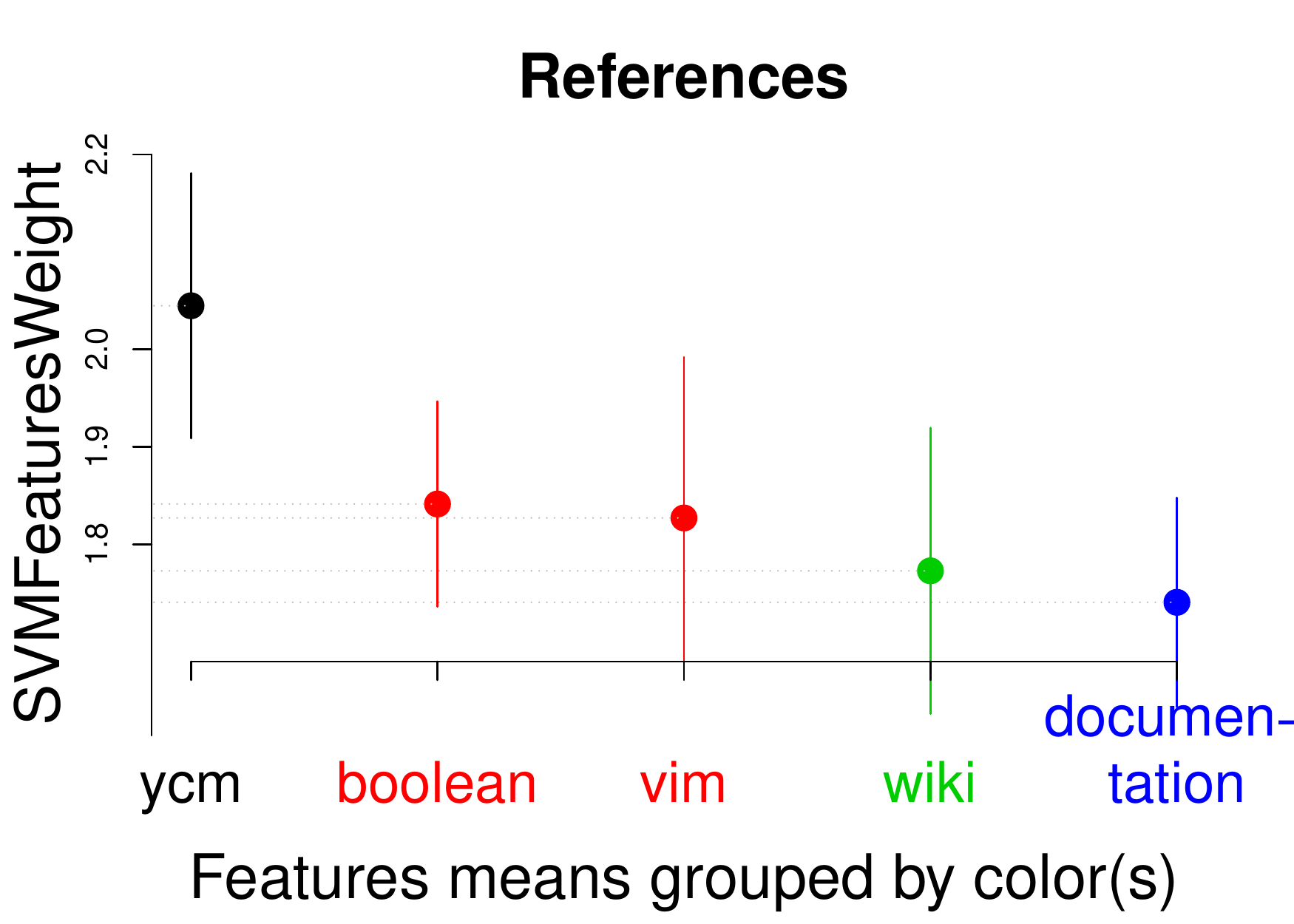}}%
\subfigure{%
\label{fig:contribution-features-rank}%
\includegraphics[width=0.5\textwidth]{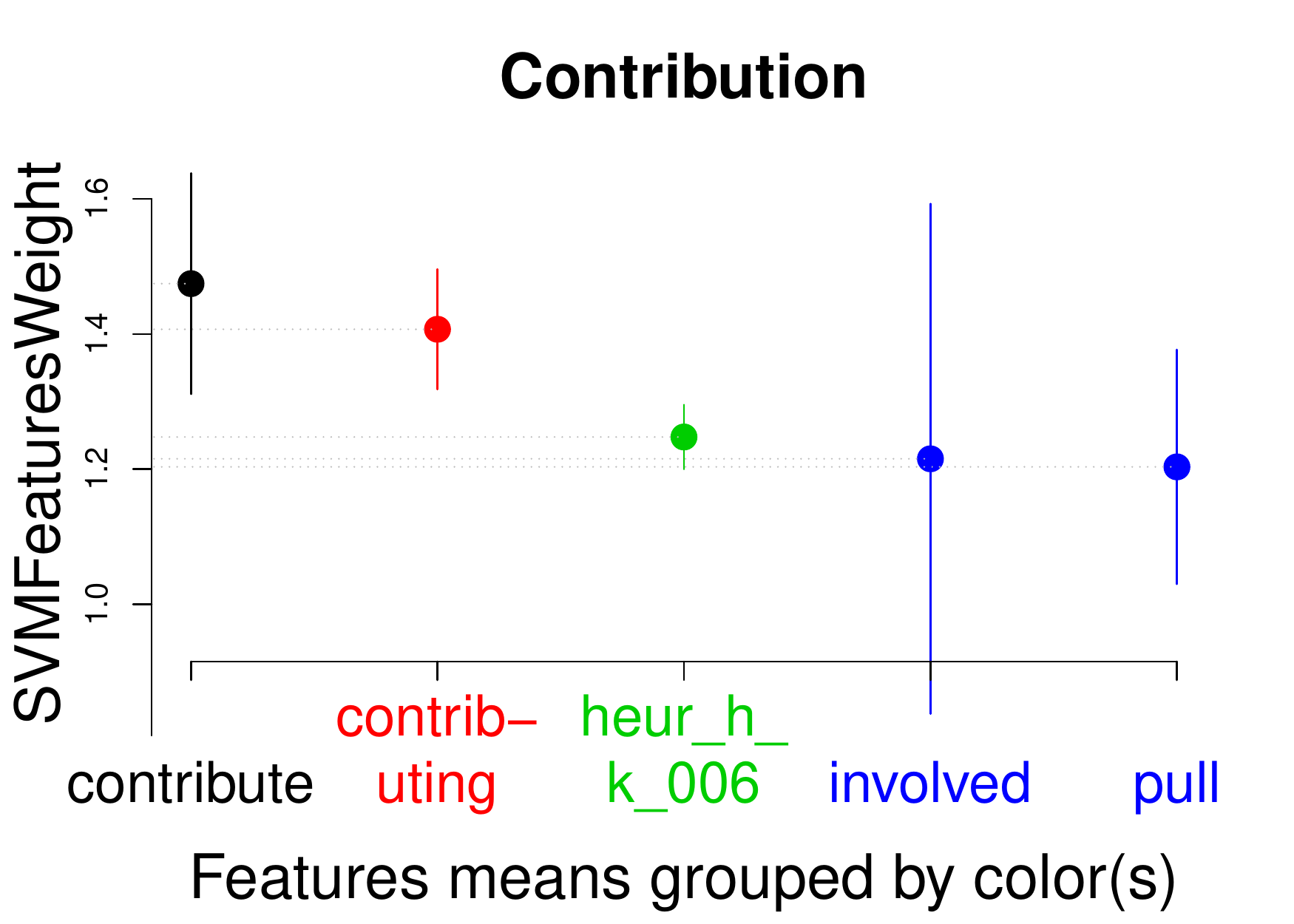}}%
\qquad
\subfigure{%
\label{fig:other-features-rank}%
\includegraphics[width=0.5\textwidth]{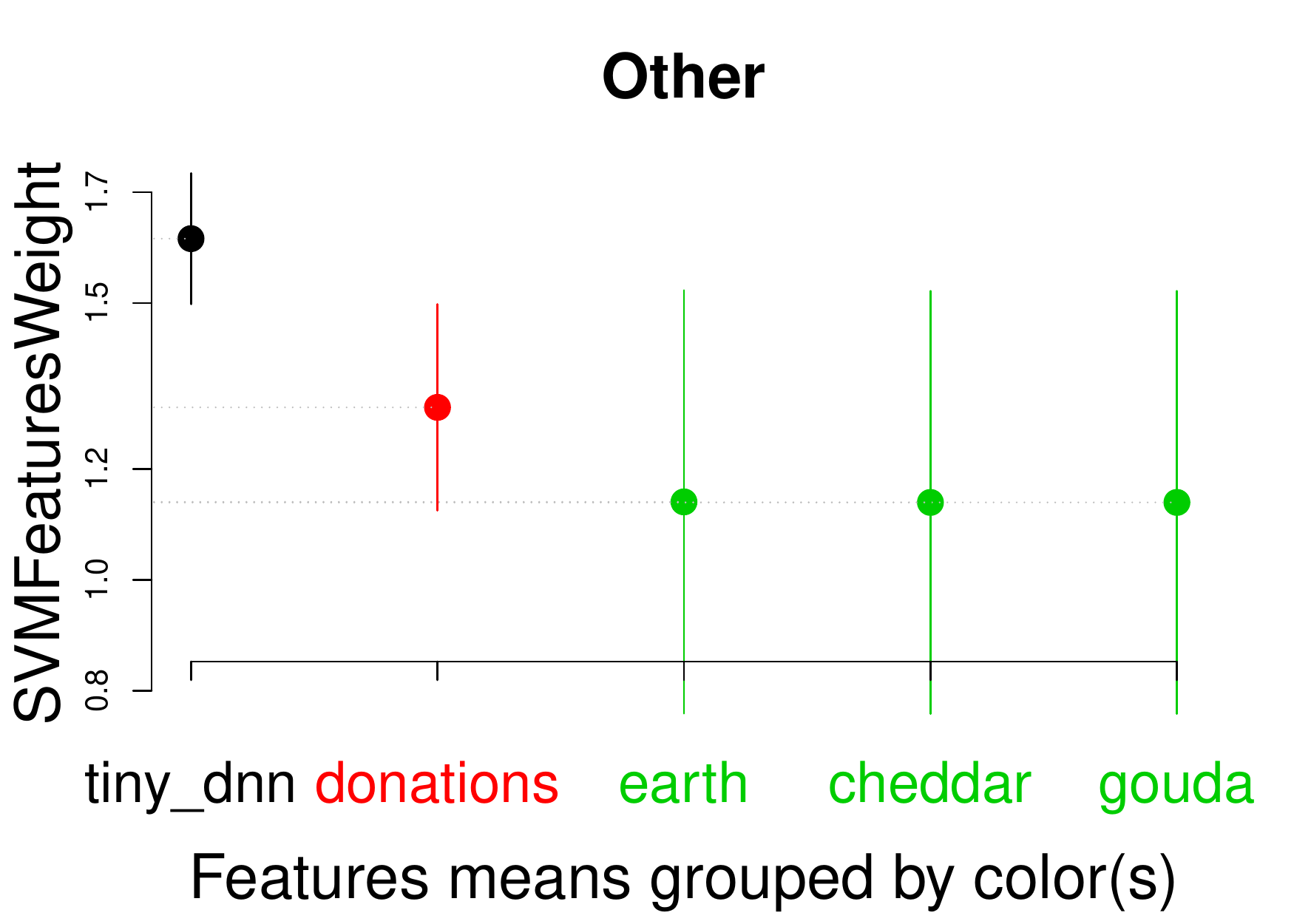}}%
\subfigure{%
\label{fig:exclusion-features-rank}%
\includegraphics[width=0.5\textwidth]{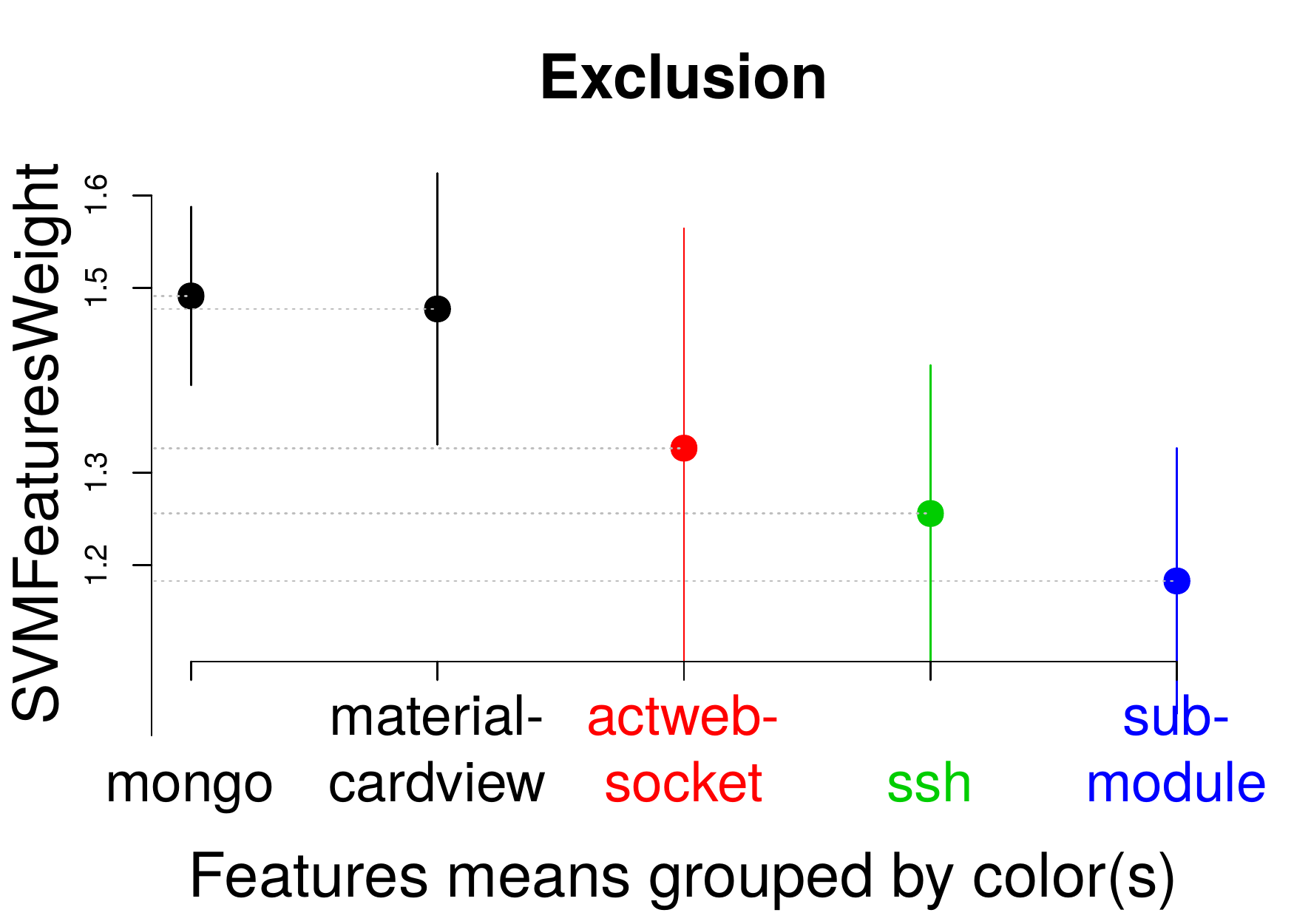}}%
\caption{Top Features for Each Category. Features starting with {\em heur\_} refer to heuristic features while the remaining features refer to statistical features (see Section 4.2).}\label{fig:top-features}
\end{figure*}

\begin{tcolorbox}
RQ3: Overall, statistical features are more useful than heuristics, but heuristics based on section headings are useful to predict certain categories
\end{tcolorbox}

\subsection{Perceived usefulness of automatically labeling sections in GitHub README files}
\label{sec:survey}

\begin{table}
\centering
\caption{Questions asked in the survey to determine perceived usefulness of automatically generated section labels}
\label{tab:survey}
\begin{tabular}{lp{10.75cm}}
\toprule
1 & Is developing software part of your job? \\
2 & What is your job title? \\
3 & For how many years have you been developing software? \\
4 & What is your area of software development? \\
5 & Do you have a GitHub account? \\
6 & Approximately how many repositories have you contributed to on GitHub? \\
7 & Have you ever contributed to the GitHub README file for a repository? \\
8 & What content do you expect to find in the README file of a GitHub repository? \\
9 & What single piece of information would you consider most important to be included in a GitHub README file? \\
10 & Is your decision to use or contribute to a GitHub project influenced by the availability of README files? \\
11 & Please take a look at the following two README files. Which one makes it easier to discover relevant information, in your opinion? Note that only the badges next to sections titles are different. \\
12 & Please justify your answer \\
13 & Do you have any further comments about GitHub README files or this survey? \\
\bottomrule
\end{tabular}
\end{table}

A potential use case for our work is to automatically label sections in GitHub README files. To evaluate the perceived usefulness of such an effort, we conducted a survey with 20 professional software developers (19 indicated to develop software as part of their job, 1 indicated to be an IT support specialist). We recruited participants using Amazon Mechanical Turk, specifying ``Employment Industry - Software \& IT Services'' as required qualification.

\begin{figure}
\centering
\includegraphics[width=\linewidth]{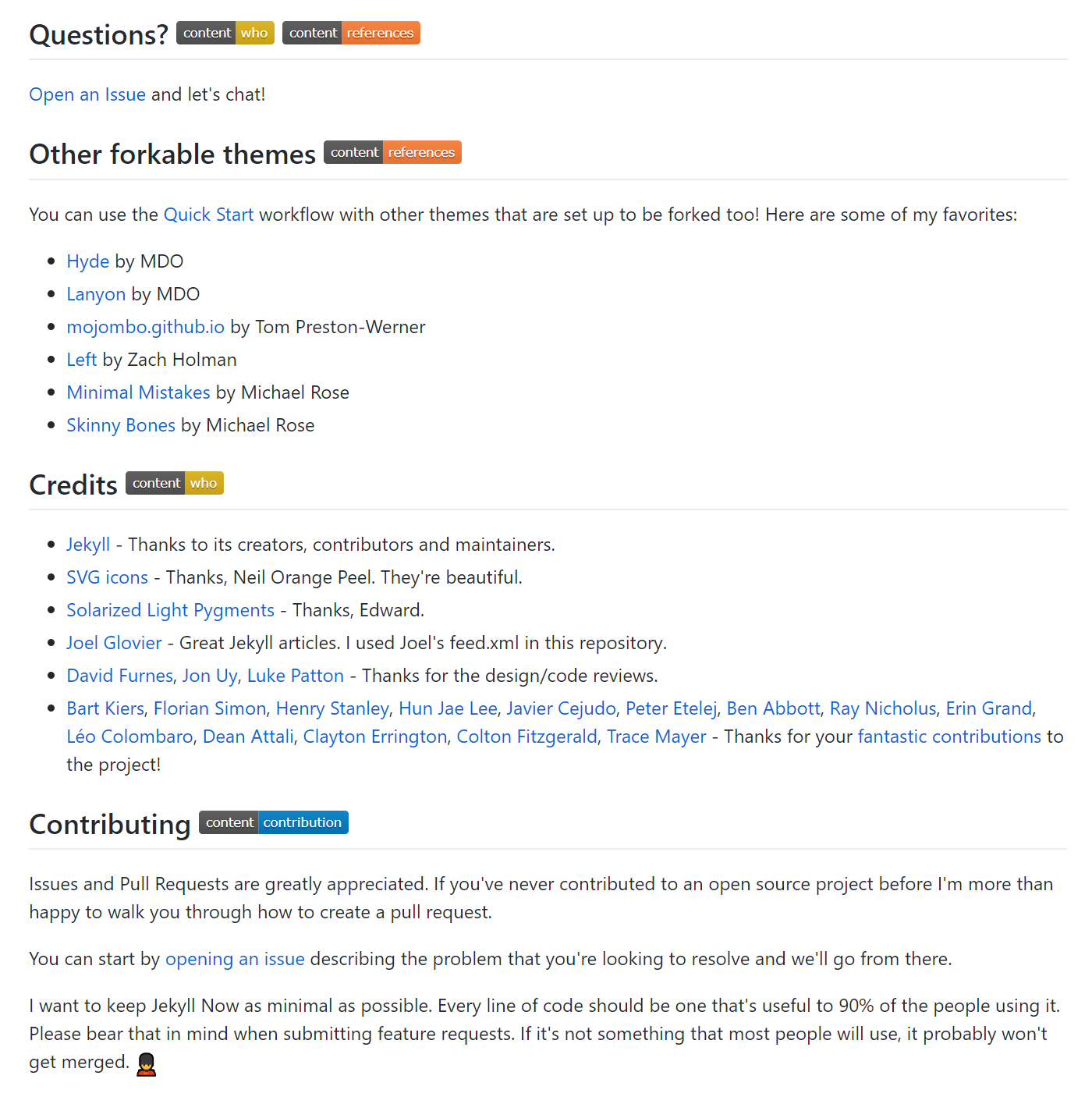}
\caption{An excerpt from a GitHub README file with visual labels, original version at \url{https://github.com/alt-blog/alt-blog.github.io}}
\label{fig:visuallabels}
\end{figure}

As part of the survey, we showed each participant two versions of a randomly selected GitHub README file which we sampled using the criteria listed in Table~\ref{tab:readme-files}. Note that the README files used for the survey were `unseen' files, i.e., files that had not been used as part of the previously introduced \textit{development} or \textit{evaluation} sets. We prepared two README files that we selected using this sampling strategy by producing two versions of each file: one version was the original README file, the other version used badges~\cite{trockman2018adding} next to each section header to indicate the labels that our classifier had automatically assigned to the section. Table~\ref{tab:survey} shows the questions asked in the survey and examples of our prepared README files are available in Figure~\ref{fig:visuallabels} and online.\footnote{Original: \url{https://github.com/readmes/alt-blog.github.io/blob/master/README1.md}, Modified: \url{https://github.com/readmes/alt-blog.github.io/blob/master/README2.md}}

All participants indicated to have been developing software for several years, with a median of five years development experience (minimum: 2 years). All but two participants indicated having a GitHub account and having contributed to more than 20 repositories on average. Only 4 of the 20 participants indicated to have never contributed to a GitHub README file.

\begin{table}
\centering
\caption{Survey results about the perceived usefulness of automatically labeled GitHub README sections}
\label{tab:survey-results}
\begin{tabular}{lr}
\toprule
prefer labeled file & 12 \\
neutral & 6 \\
prefer unlabeled file & 2 \\
\midrule
sum & 20 \\
\bottomrule
\end{tabular}
\end{table}

Table~\ref{tab:survey-results} shows the results we obtained about the perceived usefulness of the automated labeling of sections. The majority of participants (60\%) indicated that the files with our labels made it easier to discover relevant information, some participants did not have a preference, and only 2 participants preferred the unlabeled file. In general participants liked the labels, e.g., one participant wrote ``I really like the Who, what, where, and why tags. It makes it easier to find relevant information when I only need to look for a certain section.'' Similarly another participant noted: ``The what/when/how labels allow easier access to the information I am looking for.'' On the negative side, a minority of participants thought that the labels were not necessary: ``the extra buttons aren't really needed''.

\begin{tcolorbox}
RQ4: The majority of participants perceives the automated labeling of sections based on our classifier to ease information discovery in GitHub README files
\end{tcolorbox}

\section{Implications}

The ultimate goal of our work is to enable the owners of software repositories on sites such as GitHub to improve the quality of their documentation, and to make it easier for the users of the software held in these repositories to find the information they need.

The eight categories of GitHub README file content that emerged from our qualitative analysis build a point of reference for the content of such README files. These categories can help repository owners understand what content is typically included in a README file, i.e., what readers of a README file will expect to find. In this way, the categories can serve as a guideline for a README file, both for developers who are starting a new project (or who are starting the documentation for an existing project) and developers who want to evaluate the quality of their README file. Even if all the content is in place, our coding reference provides a guide on how to organize a README file.

In addition, the categories along with their frequency information that we report in this paper highlight opportunities for repository owners to stand out among a large crowd of similar repositories. For example, we found that only about a quarter of the README files in our sample contain information on the `Why' of a repository. Thus, including information on the purpose of a project is a way for repository owners to differentiate their work from that of others. It is interesting to note that out of all the kinds of content that GitHub recommends to include in a README file (cf.~Introduction), `Why' is the one that is the least represented in the README files of the repositories in our sample.

In a similar way, README content that refers to the `When' of a project, i.e., the project's current status, is rare in our sample. In order to instill confidence in its users that they are dealing with a mature software project and to possibly attract users to contribute to a project, this information is important. However, our qualitative analysis found that less than a quarter of the repositories in our random sample included `When' information.

The ratio of repositories containing information about how to contribute was slightly higher (109/393), yet surprisingly low given that all of the repositories in our sample make their source code available to the public. Given recent research on the barriers experienced by developers interested in joining open source projects~\cite{Steinmacher2016}, our findings provide another piece of evidence that software projects have room for improvement when it comes to making a good first impression~\cite{Fogel2005} and explaining how developers can contribute.

The classifier we have developed can automate the task of analyzing the content of a README file according to our coding reference, a task that would otherwise be tedious and time-consuming. Our classifier can take any GitHub README file as input and classify its content according to our codes with reasonable precision and recall. 

In addition to automatically classifying the content, our classifier could enable semi-structured access to the often unstructured information contained in a GitHub README file. For example, users particularly interested in finding mature projects could automatically be brought to the `When' sections of a README file, and developers looking to contribute to open source could be shown the `Contribution' guidelines of a repository.

The results from our survey show evidence which indicates that visually labeling sections using the labels predicted by our classifier can make it easier to find information in GitHub README files: The majority of participants perceived the automated labeling of sections based on our classifier to ease information discovery. Visually labeling sections is only one use case of the classifier: Our classifier could also easily be used to help organize README files, e.g., by imposing a certain order in which sections should appear in a README file. README sections that have been detected as discussing the `What' and `Why' of a project could automatically be moved to the beginning of a README file, followed by sections discussing the `How'. 

Our analysis of the usefulness of features for predicting the categories of a section implies that heuristic features on the sections' headings are useful, and are better suited than heuristic features on the sections' contents. This is apparent from the fact that none of the heuristic features for sections' contents are ranked among the top-5 most useful features for any of the categories. This suggests that the vocabulary commonly used in section headings is more uniform than that used in section content. However, we note that the 4,226 sections in our dataset use 3,080 distinct headings, i.e., only few of the sections share the same heading.

\section{Threats to Validity}

Similar to other empirical studies, there are several threats to the validity of our results.

Threats to the construct validity correspond to the appropriateness of the evaluation metrics. We use F1 as our evaluation metric. F1 has been used in many software engineering tasks that require classification~\cite{kim2008classifying,rahman2012recalling,nam2013transfer,canfora2013multi,rahman2013and}. Thus, we believe threats to construct validity are minimal. In our survey, we measured perceived usefulness of the visual labels added to GitHub README files, which may not correspond to actual usefulness in a software development task. Future work will have to investigate this in more detail.

Threats to the internal validity compromise our confidence in establishing a relationship between the independent and dependent variables. It is possible that we introduced bias during the manual annotation of sections from GitHub README files. We tried to mitigate this threat by using two annotators, and by manually resolving all cases in which the two annotators disagreed. We did however notice a small number of cases where annotators mistakenly treated non-sections (e.g., content that had been commented out) as sections.

Threats to external validity correspond to the ability to generalize our results. While our sample of 393 GitHub README files is statistically representative, it is plausible that a different sample of files would have generated different results. We can also not claim generalizability to any other format of software documentation. We excluded README files that were small (less than 2 KB in size), README files that belonged to repositories not used for software development, and README files not in English. Different filtering criteria might have led to different results. Our findings may also have been impacted by our decision to divide README files into sections. A different way of dividing README files (e.g., by paragraphs or sentences) might also have produced different results. Our survey was answered by twenty software professionals. We cannot claim that we have captured all possible opinions regarding the usefulness of the visual labels. All survey participants were ultimately self-selected individuals within our target
populations, and individuals who did not respond to our invitations may have different views on some of the questions that we asked. Also, creating visual labels is only one use case of our classifier, and we cannot make claims of the usefulness of other applications based on our survey results.

\section{Related Work}

Efforts related to our work can be divided into research on categories of software development knowledge, classifiers of textual content related to software engineering, and studies on the information needs of software developers.

\subsection{Categorizing software development knowledge}

Knowledge-based approaches have been extensively used in software development for decades~\cite{Ding2014}, and many research efforts have been undertaken since the 1990s to categorize the kinds of knowledge relevant to software developers~\cite{Erdos1998, Herbsleb1993, Mylopoulos1997}.

More recently, Maalej and Robillard identified 12 types of knowledge contained in API documentation, with functionality and structure being the most prevalent~\cite{Maalej2013}. Because the authors focused on API documentation, the types of knowledge they identified are more technical than ours (e.g., containing API-specific concepts such as directives), however, there is some overlap with our categorization of GitHub README files (e.g., in categories such as `References'). Similar taxonomies have been developed by Monperrus et al.~\cite{Monperrus2012} and Jeong et al.~\cite{Jeong2009}. Some of the guidelines identified by Jeong et al.~apply to our work as well (e.g., ``include `how to use' documentation'') whereas other guidelines are specific to the domain of API documentation or to the user interface through which documentation is presented (e.g., ``Effective Search''). Documentation in GitHub README files is broader than API documentation, and the documentation format and its presentation is at least partly specified by the GitHub markdown format.

In addition to API documentation, researchers have investigated the categories of knowledge contained in development blogs~\cite{Pagano2013, Parnin2011, Parnin2013, Tiarks2014} and on Stack Overflow~\cite{Asaduzzaman2013, Nasehi2012, Treude2011}. However, these formats serve different purposes compared to GitHub README files, and thus lead to different categories of software development knowledge.

\subsection{Classifying software development text}

The work most closely related to ours in terms of classifying the content of software documentation is OntoCat by Kumar and Devanbu~\cite{Kumar2016}. Using Maalej and Robillard's taxonomy of knowledge patterns in API documentation~\cite{Maalej2013}, they developed a domain independent technique to extract knowledge types from API reference documentation. Their system, OntoCat, uses nine different features and their semantic and statistical combinations to classify different knowledge types. Testing OntoCat on Python API documentation, the authors showed the effectiveness of their system. As described above, one major difference between work focused on API documentation and work on GitHub README files is that API documentation tends to be more technical. Similar to our work, Kumar and Devanbu also employed keyphrases for the classification, among other features. The F1 scores they report are in a similar range to the ones achieved by our classifier: Their weakest performance was for the categories of Non-Info (0.29) and Control Flow (0.31), while their strongest performance was for the categories of Code Examples (0.83) and Functionality and Behaviour (0.77). In our case, the lowest F1 scores were for the categories of `Other' (0.303) and `Reference' (0.605) while the highest scores were for `How' (0.861) and `Contribution' (0.814).

In other work focusing on automatically classifying the content of software documentation, Treude and Robillard developed a machine learning classifier that determines whether a sentence on Stack Overflow provides insight for a given API type~\cite{Treude2016}. Similarly, classifying content on Stack Overflow was the target of Campos et al.~\cite{Campos2014} and de Souza et al.'s work~\cite{deSouza2014}. Following on from Nasehi et al.'s categorization~\cite{Nasehi2012}, they developed classifiers to identify questions belonging to different categories, such as `How-to-do-it'. Also using data from Stack Overflow, Correa and Sureka introduced a classifier to predict deleted questions~\cite{Correa2014}.

Researchers have also applied text classification to bug reports and development issues. For example, Chaparro et al.~presented an approach to detect the absence of expected behaviour and steps to reproduce in bug descriptions, aiming to improve bug description quality by alerting reporters about missing information at reporting time~\cite{chaparro2017detecting}. Text classification has also been employed with the goal of automated generation of release notes: Moreno et al.~developed a system which extracts changes from source code, summarizes them, and integrates them with information from versioning systems and issue trackers to produce release notes~\cite{moreno2014automatic}. Abebe et al.~used machine learning techniques to automatically suggest issues to be included in release notes~\cite{Abebe2016}.

Text classification has also been applied to the information captured in other artifacts created by software developers, including change requests~\cite{Antoniol2008}, development emails~\cite{DiSorbo2015}, code comments~\cite{Pascarella2017}, requirements specifications~\cite{Mahmoud2016}, and app reviews~\cite{Chen2014, Guzman2015, Kurtanovic2017, Maalej2016}.

\subsection{Information needs of software developers}

Although there has not been much work on the information needs of software developers around GitHub repositories, there has been work on information needs of software developers in general. Early work focused mostly on program comprehension~\cite{Erdem1998, Johnson1997}. Nykaza et al.~investigated what learning support programmers need to successfully use a software development kit (SDK)~\cite{Nykaza2002}, and they catalogued the content that was seen as necessary by their interviewees, including installation instructions and documentation of system requirements. There is some overlap with the codes that emerged from our analysis, but some of Nykaza et al.'s content suggestions are SDK-specific, such as ``types of applications that can be developed with the SDK''.

Other studies on the information needs of software developers have analyzed newsgroup questions~\cite{Hou2005}, questions in collocated development teams~\cite{Ko2007, Treude2015}, questions during software evolution tasks~\cite{Sillito2006, Sillito2008}, questions that focus on issues that occur within a project~\cite{Fritz2010}, questions that are hard to answer~\cite{LaToza2010}, and information needs in software ecosystems~\cite{Haenni2013}. Information needs related to bug reports have also attracted the attention of the research community: Zimmermann et al.~conducted a survey to find out what makes a good bug report and revealed an information mismatch between what developers need and what users supply~\cite{zimmermann2010makes}. Davies and Roper investigated what information users actually provide in bug reports, how and when users provide the information, and how this affects the outcome of the bug~\cite{davies2014s}. They found that sources deemed highly useful by developers and tools such as stack traces and test cases appeared very infrequently. 

The goal of Kirk et al.'s study was understanding problems that occur during framework reuse, and they identified four problems: understanding the functionality of framework components, understanding the interactions between framework components, understanding the mapping from the problem domain to the framework implementation, and understanding the architectural assumptions in the framework design~\cite{Kirk2007}. These problems will arguably apply to frameworks hosted on GitHub, but not necessarily to other GitHub projects. Our categorization is broader by analyzing the content of GitHub README files for any type of software project. Future work might investigate README files that belong to particular kinds of projects.

\section{Conclusions and Future Work}

A README file is often the first document that a user sees when they encounter a new software repository. README files are essential in shaping the first impression of a repository and in documenting a software project. Despite their important role, we lack a systematic understanding of the content of README files as well as tools that can automate the discovery of relevant information contained in them. 

In this paper, we have reported on a qualitative study which involved the manual annotation of 4,226 sections from 393 README files for repositories hosted on GitHub. We identified eight different kinds of content, and found that information regarding the `What' and `How' of a repository is common while information on the status of a project is rare. We then designed a classifier and a set of features to automatically predict the categories of sections in README files. Our classifier achieved an F1 score of 0.746 and we found that the most useful features for classifying the content of README files were often related to particular keywords. To evaluate the usefulness of the classification, we used the automatically determined classes to label sections in GitHub README files using badges and showed files with and without these badges to twenty software professionals. The majority of participants perceived the automated labeling of sections based on our classifier to ease information discovery. 

Our findings provide a point of reference for repository owners against which they can model and evaluate their README files, ultimately leading to an improvement in the quality of software documentation. Our classifier will help automate these tasks and make it easier for users and owners of repositories to discover relevant information.

In addition to improving the precision and recall of our classifier, our future work lies in exploring the potential of the classifier to enable a more structured approach to searching and navigating GitHub README files. In particular, we plan to employ the classifier in a search interface for GitHub repositories and we will explore the feasibility of automatically reorganizing the documentation contained in GitHub README files using the structure that emerged from our qualitative analysis.

\bibliographystyle{spbasic}
\bibliography{github-readmes} 

\begin{thebibliography}{77}
\providecommand{\natexlab}[1]{#1}
\providecommand{\url}[1]{{#1}}
\providecommand{\urlprefix}{URL }
\expandafter\ifx\csname urlstyle\endcsname\relax
  \providecommand{\doi}[1]{DOI~\discretionary{}{}{}#1}\else
  \providecommand{\doi}{DOI~\discretionary{}{}{}\begingroup
  \urlstyle{rm}\Url}\fi
\providecommand{\eprint}[2][]{\url{#2}}

\bibitem[{Abebe et~al(2016)Abebe, Ali, and Hassan}]{Abebe2016}
Abebe SL, Ali N, Hassan AE (2016) An empirical study of software release notes.
  Empirical Software Engineering 21(3):1107--1142

\bibitem[{Agrawal et~al(1993)Agrawal, Imieli\'{n}ski, and Swami}]{AR}
Agrawal R, Imieli\'{n}ski T, Swami A (1993) Mining association rules between
  sets of items in large databases. In: Proceedings of the International
  Conference on Management of Data, ACM, New York, NY, USA, pp 207--216

\bibitem[{Antoniol et~al(2008)Antoniol, Ayari, Di~Penta, Khomh, and
  Gu{\'e}h{\'e}neuc}]{Antoniol2008}
Antoniol G, Ayari K, Di~Penta M, Khomh F, Gu{\'e}h{\'e}neuc YG (2008) Is it a
  bug or an enhancement?: A text-based approach to classify change requests.
  In: Proceedings of the Conference of the Center for Advanced Studies on
  Collaborative Research: Meeting of Minds, ACM, New York, NY, USA, pp
  23:304--23:318

\bibitem[{Asaduzzaman et~al(2013)Asaduzzaman, Mashiyat, Roy, and
  Schneider}]{Asaduzzaman2013}
Asaduzzaman M, Mashiyat AS, Roy CK, Schneider KA (2013) Answering questions
  about unanswered questions of {Stack} {Overflow}. In: Proceedings of the 10th
  Working Conference on Mining Software Repositories, IEEE Press, Piscataway,
  NJ, USA, pp 97--100

\bibitem[{Begel et~al(2013)Begel, Bosch, and Storey}]{Begel2013}
Begel A, Bosch J, Storey MA (2013) Social networking meets software
  development: Perspectives from {GitHub}, {MSDN}, {Stack Exchange}, and
  {TopCoder}. IEEE Software 30(1):52--66

\bibitem[{Bird et~al(2009)Bird, Klein, and Loper}]{bird2009natural}
Bird S, Klein E, Loper E (2009) Natural language processing with Python:
  analyzing text with the natural language toolkit. O'Reilly Media, Inc.

\bibitem[{Boughorbel et~al(2017)Boughorbel, Jarray, and
  El-Anbari}]{boughorbel2017optimal}
Boughorbel S, Jarray F, El-Anbari M (2017) Optimal classifier for imbalanced
  data using matthews correlation coefficient metric. PloS one 12(6):e0177,678

\bibitem[{Campos and de~Almeida~Maia(2014)}]{Campos2014}
Campos EC, de~Almeida~Maia M (2014) Automatic categorization of questions from
  {Q\&A} sites. In: Proceedings of the 29th Annual ACM Symposium on Applied
  Computing, ACM, New York, NY, USA, pp 641--643

\bibitem[{Canfora et~al(2013)Canfora, De~Lucia, Di~Penta, Oliveto, Panichella,
  and Panichella}]{canfora2013multi}
Canfora G, De~Lucia A, Di~Penta M, Oliveto R, Panichella A, Panichella S (2013)
  Multi-objective cross-project defect prediction. In: Software Testing,
  Verification and Validation (ICST), 2013 IEEE Sixth International Conference
  on, IEEE, pp 252--261

\bibitem[{Chaparro et~al(2017)Chaparro, Lu, Zampetti, Moreno, Di~Penta, Marcus,
  Bavota, and Ng}]{chaparro2017detecting}
Chaparro O, Lu J, Zampetti F, Moreno L, Di~Penta M, Marcus A, Bavota G, Ng V
  (2017) Detecting missing information in bug descriptions. In: Proceedings of
  the Joint Meeting on Foundations of Software Engineering, ACM, pp 396--407

\bibitem[{Chawla et~al(2002)Chawla, Bowyer, Hall, and
  Kegelmeyer}]{chawla2002smote}
Chawla NV, Bowyer KW, Hall LO, Kegelmeyer WP (2002) Smote: synthetic minority
  over-sampling technique. Journal of artificial intelligence research
  16:321--357

\bibitem[{Chen et~al(2014)Chen, Lin, Hoi, Xiao, and Zhang}]{Chen2014}
Chen N, Lin J, Hoi SCH, Xiao X, Zhang B (2014) Ar-miner: Mining informative
  reviews for developers from mobile app marketplace. In: Proceedings of the
  36th International Conference on Software Engineering, ACM, New York, NY,
  USA, pp 767--778

\bibitem[{Corbin and Strauss(1990)}]{Corbin1990}
Corbin JM, Strauss A (1990) Grounded theory research: Procedures, canons, and
  evaluative criteria. Qualitative Sociology 13(1):3--21

\bibitem[{Correa and Sureka(2014)}]{Correa2014}
Correa D, Sureka A (2014) Chaff from the wheat: Characterization and modeling
  of deleted questions on {Stack} {Overflow}. In: Proceedings of the 23rd
  International Conference on World Wide Web, ACM, New York, NY, USA, pp
  631--642

\bibitem[{Davies and Roper(2014)}]{davies2014s}
Davies S, Roper M (2014) What's in a bug report? In: Proceedings of the
  International Symposium on Empirical Software Engineering and Measurement,
  ACM, p~26

\bibitem[{Decan et~al(2016)Decan, Mens, Claes, and Grosjean}]{Decan2016}
Decan A, Mens T, Claes M, Grosjean P (2016) When {GitHub} meets {CRAN}: An
  analysis of inter-repository package dependency problems. In: Proceedings of
  the 23rd International Conference on Software Analysis, Evolution, and
  Reengineering, IEEE, Piscataway, NJ, USA, pp 493--504

\bibitem[{Ding et~al(2014)Ding, Liang, Tang, and Van~Vliet}]{Ding2014}
Ding W, Liang P, Tang A, Van~Vliet H (2014) Knowledge-based approaches in
  software documentation: A systematic literature review. Information and
  Software Technology 56(6):545--567

\bibitem[{Erdem et~al(1998)Erdem, Johnson, and Marsella}]{Erdem1998}
Erdem A, Johnson WL, Marsella S (1998) Task oriented software understanding.
  In: Proceedings of the 13th International Conference on Automated Software
  Engineering, IEEE Computer Society, Washington, DC, USA, pp 230--239

\bibitem[{Erd\"{o}s and Sneed(1998)}]{Erdos1998}
Erd\"{o}s K, Sneed HM (1998) Partial comprehension of complex programs (enough
  to perform maintenance). In: Proceedings of the 6th International Workshop on
  Program Comprehension, IEEE Computer Society, Washington, DC, USA, pp 98--105

\bibitem[{Fawcett(2006)}]{fawcett2006introduction}
Fawcett T (2006) An introduction to roc analysis. Pattern recognition letters
  27(8):861--874

\bibitem[{Fogel(2005)}]{Fogel2005}
Fogel K (2005) Producing Open Source Software: How to Run a Successful Free
  Software Project. O'Reilly Media, Inc., Sebastopol, CA, USA

\bibitem[{Fritz and Murphy(2010)}]{Fritz2010}
Fritz T, Murphy GC (2010) Using information fragments to answer the questions
  developers ask. In: Proceedings of the International Conference on Software
  Engineering - Volume 1, ACM, New York, NY, USA, pp 175--184

\bibitem[{Greene and Fischer(2016)}]{Greene2016}
Greene GJ, Fischer B (2016) Cvexplorer: Identifying candidate developers by
  mining and exploring their open source contributions. In: Proceedings of the
  31st IEEE/ACM International Conference on Automated Software Engineering,
  ACM, New York, NY, USA, pp 804--809

\bibitem[{Guzman et~al(2015)Guzman, El-Haliby, and Bruegge}]{Guzman2015}
Guzman E, El-Haliby M, Bruegge B (2015) Ensemble methods for app review
  classification: An approach for software evolution (n). In: Proceedings of
  the 30th International Conference on Automated Software Engineering, IEEE
  Press, Piscataway, NJ, USA, pp 771--776

\bibitem[{Haenni et~al(2013)Haenni, Lungu, Schwarz, and
  Nierstrasz}]{Haenni2013}
Haenni N, Lungu M, Schwarz N, Nierstrasz O (2013) Categorizing developer
  information needs in software ecosystems. In: Proceedings of the
  International Workshop on Ecosystem Architectures, ACM, New York, NY, USA, pp
  1--5

\bibitem[{Hassan and Wang(2017)}]{Hassan2017}
Hassan F, Wang X (2017) Mining readme files to support automatic building of
  {Java} projects in software repositories: Poster. In: Proceedings of the 39th
  International Conference on Software Engineering Companion, IEEE Press,
  Piscataway, NJ, USA, pp 277--279

\bibitem[{Hauff and Gousios(2015)}]{Hauff2015}
Hauff C, Gousios G (2015) Matching {GitHub} developer profiles to job
  advertisements. In: Proceedings of the 12th Working Conference on Mining
  Software Repositories, IEEE Press, Piscataway, NJ, USA, pp 362--366

\bibitem[{Herbsleb and Kuwana(1993)}]{Herbsleb1993}
Herbsleb JD, Kuwana E (1993) Preserving knowledge in design projects: What
  designers need to know. In: Proceedings of the INTERACT '93 and CHI '93
  Conference on Human Factors in Computing Systems, ACM, New York, NY, USA, pp
  7--14

\bibitem[{Hou et~al(2005)Hou, Wong, and Hoover}]{Hou2005}
Hou D, Wong K, Hoover HJ (2005) What can programmer questions tell us about
  frameworks? In: Proceedings of the 13th International Workshop on Program
  Comprehension, IEEE, Piscataway, NJ, USA, pp 87--96

\bibitem[{Jeong et~al(2009)Jeong, Xie, Beaton, Myers, Stylos, Ehret, Karstens,
  Efeoglu, and Busse}]{Jeong2009}
Jeong SY, Xie Y, Beaton J, Myers BA, Stylos J, Ehret R, Karstens J, Efeoglu A,
  Busse DK (2009) Improving documentation for {eSOA} {APIs} through user
  studies. In: Proceedings of the 2nd International Symposium on End-User
  Development, Springer-Verlag, Berlin, Heidelberg, pp 86--105

\bibitem[{Johnson and Erdem(1997)}]{Johnson1997}
Johnson WL, Erdem A (1997) Interactive explanation of software systems.
  Automated Software Engineering 4(1):53--75

\bibitem[{Kalliamvakou et~al(2014)Kalliamvakou, Gousios, Blincoe, Singer,
  German, and Damian}]{Kalliamvakou2014}
Kalliamvakou E, Gousios G, Blincoe K, Singer L, German DM, Damian D (2014) The
  promises and perils of mining {GitHub}. In: Proceedings of the 11th Working
  Conference on Mining Software Repositories, ACM, New York, NY, USA, pp
  92--101

\bibitem[{Kim et~al(2008)Kim, Whitehead~Jr, and Zhang}]{kim2008classifying}
Kim S, Whitehead~Jr EJ, Zhang Y (2008) Classifying software changes: Clean or
  buggy? IEEE Transactions on Software Engineering 34(2):181--196

\bibitem[{Kirk et~al(2007)Kirk, Roper, and Wood}]{Kirk2007}
Kirk D, Roper M, Wood M (2007) Identifying and addressing problems in
  object-oriented framework reuse. Empirical Software Engineering
  12(3):243--274

\bibitem[{Ko et~al(2007)Ko, DeLine, and Venolia}]{Ko2007}
Ko AJ, DeLine R, Venolia G (2007) Information needs in collocated software
  development teams. In: Proceedings of the 29th International Conference on
  Software Engineering, IEEE Computer Society, Washington, DC, USA, pp 344--353

\bibitem[{Kumar and Devanbu(2016)}]{Kumar2016}
Kumar N, Devanbu PT (2016) Ontocat: Automatically categorizing knowledge in
  {API} documentation. CoRR abs/1607.07602:preprint

\bibitem[{Kurtanovi\'{c} and Maalej(2017)}]{Kurtanovic2017}
Kurtanovi\'{c} Z, Maalej W (2017) Mining user rationale from software reviews.
  In: Proceedings of the 25th International Requirements Engineering
  Conference, IEEE, Piscataway, NJ, USA, pp 61--70

\bibitem[{Landis and Koch(1977)}]{landis1977measurement}
Landis JR, Koch GG (1977) The measurement of observer agreement for categorical
  data. biometrics pp 159--174

\bibitem[{LaToza and Myers(2010)}]{LaToza2010}
LaToza TD, Myers BA (2010) Hard-to-answer questions about code. In: Evaluation
  and Usability of Programming Languages and Tools, ACM, New York, NY, USA, pp
  8:1--8:6

\bibitem[{Lessmann et~al(2008)Lessmann, Baesens, Mues, and
  Pietsch}]{lessmann2008benchmarking}
Lessmann S, Baesens B, Mues C, Pietsch S (2008) Benchmarking classification
  models for software defect prediction: A proposed framework and novel
  findings. IEEE Transactions on Software Engineering 34(4):485--496

\bibitem[{Luaces et~al(2012)Luaces, D{\'\i}ez, Barranquero, del Coz, and
  Bahamonde}]{luaces2012binary}
Luaces O, D{\'\i}ez J, Barranquero J, del Coz JJ, Bahamonde A (2012) Binary
  relevance efficacy for multilabel classification. Progress in Artificial
  Intelligence 1(4):303--313

\bibitem[{Maalej and Robillard(2013)}]{Maalej2013}
Maalej W, Robillard MP (2013) Patterns of knowledge in {API} reference
  documentation. IEEE Transactions on Software Engineering 39(9):1264--1282

\bibitem[{Maalej et~al(2016)Maalej, Kurtanovi\'{c}, Nabil, and
  Stanik}]{Maalej2016}
Maalej W, Kurtanovi\'{c} Z, Nabil H, Stanik C (2016) On the automatic
  classification of app reviews. Requirements Engineering 21(3):311--331

\bibitem[{Mahmoud and Williams(2016)}]{Mahmoud2016}
Mahmoud A, Williams G (2016) Detecting, classifying, and tracing non-functional
  software requirements. Requirements Engineering 21(3):357--381

\bibitem[{Miles and Huberman(1994)}]{Miles1994}
Miles MB, Huberman AM (1994) Qualitative Data Analysis: An Expanded Sourcebook.
  SAGE publications

\bibitem[{Monperrus et~al(2012)Monperrus, Eichberg, Tekes, and
  Mezini}]{Monperrus2012}
Monperrus M, Eichberg M, Tekes E, Mezini M (2012) What should developers be
  aware of? an empirical study on the directives of api documentation.
  Empirical Software Engineering 17(6):703--737

\bibitem[{Moreno et~al(2014)Moreno, Bavota, Di~Penta, Oliveto, Marcus, and
  Canfora}]{moreno2014automatic}
Moreno L, Bavota G, Di~Penta M, Oliveto R, Marcus A, Canfora G (2014) Automatic
  generation of release notes. In: Proceedings of the International Symposium
  on Foundations of Software Engineering, ACM, pp 484--495

\bibitem[{Mylopoulos et~al(1997)Mylopoulos, Borgida, and Yu}]{Mylopoulos1997}
Mylopoulos J, Borgida A, Yu E (1997) Representing software engineering
  knowledge. Automated Software Engineering 4(3):291--317

\bibitem[{Nam et~al(2013)Nam, Pan, and Kim}]{nam2013transfer}
Nam J, Pan SJ, Kim S (2013) Transfer defect learning. In: Proceedings of the
  2013 International Conference on Software Engineering, IEEE Press, pp
  382--391

\bibitem[{Nasehi et~al(2012)Nasehi, Sillito, Maurer, and Burns}]{Nasehi2012}
Nasehi SM, Sillito J, Maurer F, Burns C (2012) What makes a good code example?:
  A study of programming {Q\&A} in {StackOverflow}. In: Proceedings of the
  International Conference on Software Maintenance, IEEE Computer Society,
  Washington, DC, USA, pp 25--34

\bibitem[{Nykaza et~al(2002)Nykaza, Messinger, Boehme, Norman, Mace, and
  Gordon}]{Nykaza2002}
Nykaza J, Messinger R, Boehme F, Norman CL, Mace M, Gordon M (2002) What
  programmers really want: Results of a needs assessment for sdk documentation.
  In: Proceedings of the 20th Annual International Conference on Computer
  Documentation, ACM, New York, NY, USA, pp 133--141

\bibitem[{Pagano and Maalej(2013)}]{Pagano2013}
Pagano D, Maalej W (2013) How do open source communities blog? Empirical
  Software Engineering 18(6):1090--1124

\bibitem[{Panichella et~al(2015)Panichella, Di~Sorbo, Guzman, Visaggio,
  Canfora, and Gall}]{panichella2015can}
Panichella S, Di~Sorbo A, Guzman E, Visaggio CA, Canfora G, Gall HC (2015) How
  can i improve my app? classifying user reviews for software maintenance and
  evolution. In: Software maintenance and evolution (ICSME), 2015 IEEE
  international conference on, IEEE, pp 281--290

\bibitem[{Parnin and Treude(2011)}]{Parnin2011}
Parnin C, Treude C (2011) Measuring {API} documentation on the web. In:
  Proceedings of the 2nd International Workshop on Web 2.0 for Software
  Engineering, ACM, New York, NY, USA, pp 25--30

\bibitem[{Parnin et~al(2013)Parnin, Treude, and Storey}]{Parnin2013}
Parnin C, Treude C, Storey MA (2013) Blogging developer knowledge: Motivations,
  challenges, and future directions. In: Proceedings of the 21st International
  Conference on Program Comprehension, IEEE Press, Piscataway, NJ, USA, pp
  211--214

\bibitem[{Pascarella and Bacchelli(2017)}]{Pascarella2017}
Pascarella L, Bacchelli A (2017) Classifying code comments in java open-source
  software systems. In: Proceedings of the 14th International Conference on
  Mining Software Repositories, IEEE Press, Piscataway, NJ, USA, pp 227--237

\bibitem[{Pedregosa et~al(2011)Pedregosa, Varoquaux, Gramfort, Michel, Thirion,
  Grisel, Blondel, Prettenhofer, Weiss, Dubourg et~al}]{pedregosa2011scikit}
Pedregosa F, Varoquaux G, Gramfort A, Michel V, Thirion B, Grisel O, Blondel M,
  Prettenhofer P, Weiss R, Dubourg V, et~al (2011) Scikit-learn: Machine
  learning in python. Journal of Machine Learning Research 12(Oct):2825--2830

\bibitem[{Portugal and do~Prado~Leite(2016)}]{Portugal2016}
Portugal RLQ, do~Prado~Leite JCS (2016) Extracting requirements patterns from
  software repositories. In: Proceedings of the 24th International Requirements
  Engineering Conference Workshops, IEEE, Piscataway, NJ, USA, pp 304--307

\bibitem[{Prasetyo et~al(2012)Prasetyo, Lo, Achananuparp, Tian, and
  Lim}]{prasetyo2012automatic}
Prasetyo PK, Lo D, Achananuparp P, Tian Y, Lim EP (2012) Automatic
  classification of software related microblogs. In: Software Maintenance
  (ICSM), 2012 28th IEEE International Conference on, IEEE, pp 596--599

\bibitem[{Rahman and Devanbu(2013)}]{rahman2013and}
Rahman F, Devanbu P (2013) How, and why, process metrics are better. In:
  Proceedings of the 2013 International Conference on Software Engineering,
  IEEE Press, pp 432--441

\bibitem[{Rahman et~al(2012)Rahman, Posnett, and Devanbu}]{rahman2012recalling}
Rahman F, Posnett D, Devanbu P (2012) Recalling the imprecision of
  cross-project defect prediction. In: Proceedings of the ACM SIGSOFT 20th
  International Symposium on the Foundations of Software Engineering, ACM, pp
  61:1--61:11

\bibitem[{Romano and Pinzger(2011)}]{romano2011using}
Romano D, Pinzger M (2011) Using source code metrics to predict change-prone
  java interfaces. In: Software Maintenance (ICSM), 2011 27th IEEE
  International Conference on, IEEE, pp 303--312

\bibitem[{Sharma et~al(2017)Sharma, Thung, Kochhar, Sulistya, and
  Lo}]{Sharma2017}
Sharma A, Thung F, Kochhar PS, Sulistya A, Lo D (2017) Cataloging {GitHub}
  repositories. In: Proceedings of the 21st International Conference on
  Evaluation and Assessment in Software Engineering, ACM, New York, NY, USA, pp
  314--319

\bibitem[{Sillito et~al(2006)Sillito, Murphy, and De~Volder}]{Sillito2006}
Sillito J, Murphy GC, De~Volder K (2006) Questions programmers ask during
  software evolution tasks. In: Proceedings of the International Symposium on
  the Foundations of Software Engineering, ACM, New York, NY, USA, pp 23--34

\bibitem[{Sillito et~al(2008)Sillito, Murphy, and De~Volder}]{Sillito2008}
Sillito J, Murphy GC, De~Volder K (2008) Asking and answering questions during
  a programming change task. IEEE Transactions on Software Engineering
  34(4):434--451

\bibitem[{Sorbo et~al(2015)Sorbo, Panichella, Visaggio, Penta, Canfora, and
  Gall}]{DiSorbo2015}
Sorbo AD, Panichella S, Visaggio CA, Penta MD, Canfora G, Gall HC (2015)
  Development emails content analyzer: Intention mining in developer
  discussions (t). In: Proceedings of the 30th International Conference on
  Automated Software Engineering, IEEE Press, Piscataway, NJ, USA, pp 12--23

\bibitem[{de~Souza et~al(2014)de~Souza, Campos, and Maia}]{deSouza2014}
de~Souza LBL, Campos EC, Maia MdA (2014) Ranking crowd knowledge to assist
  software development. In: Proceedings of the 22nd International Conference on
  Program Comprehension, ACM, New York, NY, USA, pp 72--82

\bibitem[{Steinmacher et~al(2016)Steinmacher, Conte, Treude, and
  Gerosa}]{Steinmacher2016}
Steinmacher I, Conte TU, Treude C, Gerosa MA (2016) Overcoming open source
  project entry barriers with a portal for newcomers. In: Proceedings of the
  38th International Conference on Software Engineering, ACM, New York, NY,
  USA, pp 273--284

\bibitem[{Tantithamthavorn et~al(2017)Tantithamthavorn, McIntosh, Hassan, and
  Matsumoto}]{tantithamthavorn2017mvt}
Tantithamthavorn C, McIntosh S, Hassan AE, Matsumoto K (2017) An empirical
  comparison of model validation techniques for defect prediction models. IEEE
  Transactions on Software Engineering 43(1):1--18

\bibitem[{Tiarks and Maalej(2014)}]{Tiarks2014}
Tiarks R, Maalej W (2014) How does a typical tutorial for mobile development
  look like? In: Proceedings of the 11th Working Conference on Mining Software
  Repositories, ACM, New York, NY, USA, pp 272--281

\bibitem[{Treude and Robillard(2016)}]{Treude2016}
Treude C, Robillard MP (2016) Augmenting {API} documentation with insights from
  {Stack} {Overflow}. In: Proceedings of the 38th International Conference on
  Software Engineering, ACM, New York, NY, USA, pp 392--403

\bibitem[{Treude et~al(2011)Treude, Barzilay, and Storey}]{Treude2011}
Treude C, Barzilay O, Storey MA (2011) How do programmers ask and answer
  questions on the web? ({NIER} track). In: Proceedings of the 33rd
  International Conference on Software Engineering, ACM, New York, NY, USA, pp
  804--807

\bibitem[{Treude et~al(2015)Treude, Figueira~Filho, and Kulesza}]{Treude2015}
Treude C, Figueira~Filho F, Kulesza U (2015) Summarizing and measuring
  development activity. In: Proceedings of the 10th Joint Meeting on
  Foundations of Software Engineering, ACM, New York, NY, USA, pp 625--636

\bibitem[{Trockman et~al(2018)Trockman, Zhou, K{\"a}stner, and
  Vasilescu}]{trockman2018adding}
Trockman A, Zhou S, K{\"a}stner C, Vasilescu B (2018) Adding sparkle to social
  coding: an empirical study of repository badges in the npm ecosystem. In:
  Proceedings of the 40th International Conference on Software Engineering,
  ACM, pp 511--522

\bibitem[{Xia et~al(2014)Xia, Feng, Lo, Chen, and Wang}]{xia2014towards}
Xia X, Feng Y, Lo D, Chen Z, Wang X (2014) Towards more accurate multi-label
  software behavior learning. In: Software Maintenance, Reengineering and
  Reverse Engineering (CSMR-WCRE), 2014 Software Evolution Week-IEEE Conference
  on, IEEE, pp 134--143

\bibitem[{Zhang et~al(2017)Zhang, Lo, Kochhar, Xia, Li, and Sun}]{Zhang2017}
Zhang Y, Lo D, Kochhar PS, Xia X, Li Q, Sun J (2017) Detecting similar
  repositories on {GitHub}. In: Proceedings of the 24th International
  Conference on Software Analysis, Evolution and Reengineering, IEEE,
  Piscataway, NJ, USA, pp 13--23

\bibitem[{Zimmermann et~al(2010)Zimmermann, Premraj, Bettenburg, Just,
  Schr\"{o}ter, and Weiss}]{zimmermann2010makes}
Zimmermann T, Premraj R, Bettenburg N, Just S, Schr\"{o}ter A, Weiss C (2010)
  What makes a good bug report? IEEE Transactions on Software Engineering
  36(5):618--643

\end{thebibliography}
\end{sloppy}
\end{document}